\documentclass[          
 pre,
 twocolumn,
 twoside,
 superscriptaddress,
 amsmath,
 amssymb,                
 apsf
]{revtex4-2}

\usepackage[english]{babel}
\usepackage{float}
\usepackage{graphicx}
\usepackage{tikz}
\usepackage{booktabs}
\usepackage{subcaption} 
\usetikzlibrary{arrows.meta, positioning}

\definecolor{accent1}{HTML}{7CA982}
\definecolor{accent2}{HTML}{DB9D47}
\definecolor{todoblue}{RGB}{0, 91, 187}

\usepackage{hyperref}
\hypersetup{
  hypertexnames=false,
  colorlinks=true,
  allcolors=todoblue,
  urlcolor=.,
  linkcolor=todoblue,
  citecolor=todoblue,
  pdfborder={0 0 0},
  breaklinks=true,
}

\definecolor{darkbg}{RGB}{15, 18, 25}
\definecolor{lighttext}{RGB}{247, 247, 255}
\definecolor{accent}{RGB}{211, 63, 73}

\usepackage{caption}
\usepackage{subcaption}

\makeatletter
\let\ftype@table\ftype@figure
\makeatother

\begin{document}

%% \title{``Misinformation'' before and after 2016: From the Satanic panic to social media}

\title{False memories to fake news:
\\
The evolution of the term ``misinformation'' in  
academic literature 
}

%% \\
%% from the Satanic panic to mainstream politics}

% Distrust, conspiracism, and virality: How ``misinformation'' transferred from the Satanic panic to mainstream politics
% ``Misinformation'' before and after 2016: the transition from the Satanic Panic and false memories to platforms and Fake News
% ``Misinformation'' before and after 2016: from false memories to fake news
% ``Misinformation'' from false memories to fake news: how distrust, conspiracism, and virality underpin the Satanic panic and contemporary politics
% ``Misinformation'' from false memories to fake news: the evolution of the academic term from the Satanic panic to mainstream politics
% ``Misinformation'' from false memories to fake news: the evolution of the term in the academic (scientific?) literature from the Satanic panic to mainstream politics

%%%%%%%%%%%%%%%%%%%%%%%%%%%%%
%% author list
%%%%%%%%%%%%%%%%%%%%%%%%%%%%%

\author{
\firstname{Alejandro}
\surname{Javier Ruiz Iglesias}
}

\email{alejandro.ruiz@uvm.edu}

\affiliation{
  Computational Story Lab,
  Vermont Complex Systems Institute,
  MassMutual Center of Excellence in Complex Systems and Data Science,
  University of Vermont,
  Burlington, VT 05405, USA.
  }

\author{
\firstname{Danny}
\surname{Benett}
}

\affiliation{
  Computational Story Lab,
  Vermont Complex Systems Institute,
  MassMutual Center of Excellence in Complex Systems and Data Science,
  University of Vermont,
  Burlington, VT 05405, USA.
  }

\author{
\firstname{Julia Witte}
\surname{Zimmerman}
}

\affiliation{
  Computational Story Lab,
  Vermont Complex Systems Institute,
  MassMutual Center of Excellence in Complex Systems and Data Science,
  University of Vermont,
  Burlington, VT 05405, USA.
  }

\author{
\firstname{Christopher M.}
\surname{Danforth}
}

\affiliation{
  Computational Story Lab,
  Vermont Complex Systems Institute,
  MassMutual Center of Excellence in Complex Systems and Data Science,
  University of Vermont,
  Burlington, VT 05405, USA.
  }

\affiliation{
  Department of Mathematics \& Statistics,
  University of Vermont,
  Burlington, VT 05405, USA.
  }

\author{
  \firstname{Peter Sheridan}
  \surname{Dodds}
}

\email{peter.dodds@uvm.edu}

\affiliation{
  Computational Story Lab,
  Vermont Complex Systems Institute,
  MassMutual Center of Excellence in Complex Systems and Data Science,
  University of Vermont,
  Burlington, VT 05405, USA.
  }

\affiliation{
  Department of Computer Science,
  University of Vermont,
  Burlington, VT 05405, USA.
}

\affiliation{
  Santa Fe Institute,
  1399 Hyde Park Rd,
  Santa Fe,
  NM 87501,
  USA
}

\affiliation{
  Complexity Science Hub,
  Metternichgasse 8,
  1030 Vienna,
  Austria
}

\begin{abstract}
   Since 2016, the term ``misinformation'' has become associated with a scientific paradigm that studies, at its core, people making, reading, and sharing false statements, usually on social media, and often warning of the harm to society resulting from the sum of many such events. By tracking the term through the academic literature, with special focus on the years 2011--2023, we connect this post-2016 paradigm with a strand of research dating to the Satanic panic of the 1980s. We argue that post-2016 misinformation research owes more to this intellectual lineage than is generally acknowledged, and we discuss the theoretical and practical implications of this connection. We conclude by drawing parallels between the Satanic panic and 2026, and, similarly, between misinformation research then and now. 
\end{abstract}

\maketitle

\section{Introduction}
\label{intro}

In the academic literature, the term ``misinformation''  has come to represent a new scientific paradigm. Like any paradigm, it comes with its own history and assumptions, and has brought together distinct strands of research under a shared framework and common language~\cite{kuhn_structure_1962}, which includes related terms like ``disinformation,'' ``fake news,'' and ``infodemic.'' As always, with a new paradigm's success comes a re-interpretation of previously-known facts. Misinformation has, for example, been used to describe events in Ancient Rome~\cite{posetti_short_2018}, and even the activities of fish and other animals~\cite{fahimipour_wild_2023, kongBriefNaturalHistory2025, norburyMisinformationTacticsProtect2021}. The paradigm's popularity means that such studies can garner attention in the popular press~\cite{zimmer_for_2025,arnoldMisinformationInevitableBiological}.

\begin{figure}[tp]
\centering
\includegraphics[width=\linewidth]{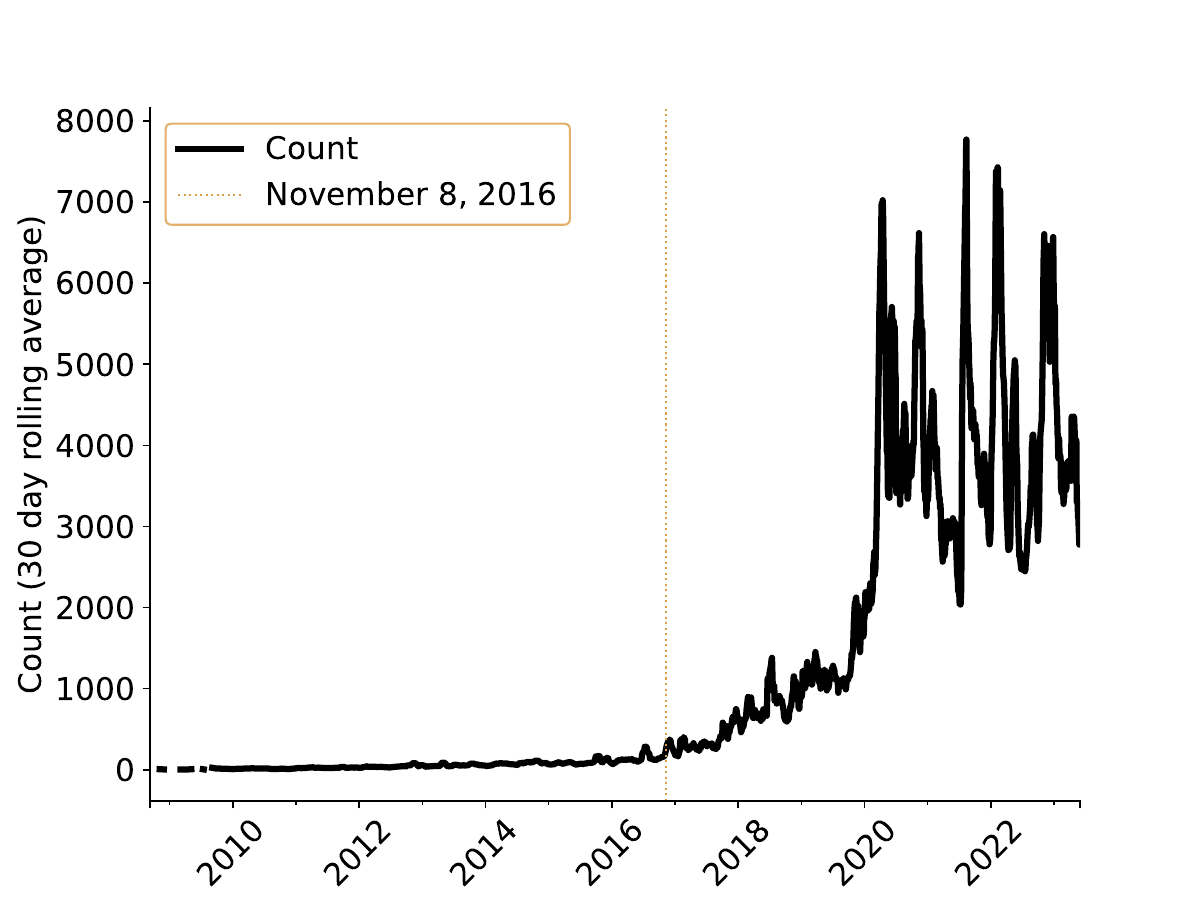}
\caption{Count of the word ``misinformation'' in ~\citet{alshaabi2021storywrangler}'s Twitter data. The date of the 2016 US Presidential Election appears to roughly coincide with the beginning of a new trend of growth.}
\label{fig:twitter}
\end{figure}

Though it has none of the paradigm-defining weight it carries in 2026, the term can be found in the academic literature of the 1950s. At the time, it was a relatively uncommon word, almost entirely used in medicine or public health. ``Combating Food Misinformation and Quackery''~\cite{HUENEMANN1956623}, for example, contains a familiar use of the term, and dates back to 1956. In the next year, however, we find ``Information and Misinformation Gained from Fasting Blood Sugar Alone in Diabetes Therapy''~\cite{johnInformationMisinformationGained1957}.

This trend continues through the 1960s, with similar usage about the quality of information gained from some process, social or otherwise, but now across increasingly diverse fields. Public health and medicine still dominate, but one finds, for example, ``A Model of the Soviet Firm''~\cite{gindinModelSovietFirm1970}, which discusses Soviet firms' attempts to correct for ``systematic misinformation'' by changing the incentives of compensation structure. In the 1970s through the 1990s, we start to see more specialized uses, with perhaps slightly divergent, more technical (if metaphorically-linked) meanings, like those associated with computation~\cite{michelCategoricalApproachDistributed1989}.

Among these usages comes that of the ``misinformation effect,'' in which ``[e]xposure to misinformation after an event takes place puts memory accuracy at risk''~\cite{loftus_misinformation_2024}. In \citet{loftus_reconstruction_1974}, we see the beginning of a paradigm that will identify itself closely with the term. Its experiments might involve asking a subject to recall something and, in the process, attempting to influence that memory in some way, perhaps by implication in the phrasing of a 
question~\cite{loftusPlantingMisinformationHuman2005b}. In a striking historical parallel to the 2020s, this misinformation paradigm played an important role in public life during the Satanic panic of the 1980s and 1990s. We will discuss this usage at some length in Sections~\ref{results} and \ref{discussion}. 

In the early 2000s, public health practitioners become interested in ``infodemiology'', or ``the study of the determinants and distribution of health information and misinformation—which may be useful in guiding health professionals and patients to quality health information on the Internet''~\cite{eysenbachInfodemiologyEpidemiologyMisinformation2002}. From there, in what seems a logical next step, we see papers that study how (mis)information spreads during crises on social media, including epidemics~\cite{suttonBackchannelsFrontLines2008, chewPandemicsAgeTwitter2010, ohEXPLORATIONSOCIALMEDIA2010}.

Finally, in the 2020s, as we show in Section~\ref{results}, we have a new usage of misinformation, one focused on social media, and one that has entered into popular consciousness (e.g., see Figure~\ref{fig:twitter}). It is not new that people say things that are not true, whether on social media or otherwise. That people often believe these things, both to our individual and collective detriments, is similarly unsurprising. In the 2020s, however, in both formal and informal settings, we have described this phenomenon as misinformation in a specific yet broadly applicable sense. 

Previous work shows, either implicitly or explicitly, that the paradigm dates to roughly 2016, and that its existence is intertwined with the rise of social media~\cite{broda_misinformation_2024,donovan_express_2025}. Some argue this point in a critique of the paradigm~\cite{miro-llinares_misinformation_2023}. These histories tend to begin with a general statement on the vast history of people saying and believing untrue things. They often cite technological upheavals in media as important factors, then specifically turn their focus to social media. 

Here, by analyzing academic papers published between 2011 and 2023 with the keyword ``misinformation'' in their metadata, we identify pre-2016 linguistic patterns unaccounted for in these histories~\cite{broda_misinformation_2024,donovan_express_2025}, and we connect these patterns to the research on the misinformation effect. We argue that post-2016 misinformation research owes more to this lineage than is generally acknowledged. 

We conclude by noting the parallels between the misinformation effect research of the 1980s and that of the 2020s. Both came about in moments of mass panic about evil forces doing harm to children, fueled by a changing media and political context. We are by no means the first to note the historical parallels between Qanon and the Satanic panic, and we are much indebted to prior scholarship and journalism on this issue. Instead, our contribution is to note that, just as~\citet{hearst_qanon_2022} finds that modern conspiricism has roots in the Satanic panic, so too can we trace the modern scientific response back to that of the 1980s. To many of the field's most cited authors, especially those working in the tradition of the misinformation effect, this history is presumably well-known. Instead, we hope that, by making this history more explicit, we can add context to existing histories and literature reviews, and that, with this context, the scientific community can better navigate its own cultural authority in these moments of epistemic struggle.

\section{Data and Methods}

Our primary database is a Scopus~\cite{scopus} export of all papers with ``misinformation'' in any metadata field from 2011–2023. We supplemented that with the opencitations.net API.

\subsection{Term Frequency}

We split our corpus into two case-insensitive bags of words (BOWs): One using all titles and abstracts from the years 2011-2015, and the second from 2017 onward. We chose to split on the year 2016 because it is the year when the term's sudden surge in popularity began. We exclude 2016 from either BOW as a transition year because, had we added it to pre-2016, the beginning of the surge dominates the behavior in an already imbalanced comparison. Though it would affect the analysis less, we also see evidence against adding it to post-2016 side because the transition is still underway. Figure~\ref{fig:twitter}, for example, shows the usage spike towards the end of 2016. Given that scientific publishing takes time, we opt instead to exclude the year entirely.

When tokenizing, we kept bigrams that occur together more than 100 times as a single token. We stripped punctuation, the most common stopwords, and some stopwords specific to our corpus (e.g., "elsevier" and "llc"). We calculated the frequency of every token in both BOWs, then chose the 9 smallest (i.e., most negative) and largest differences of frequency from the first to the second. We then classified all papers as having at least one word from either set or one of both.

\begin{figure}[t]
\centering
\includegraphics[width=\linewidth]{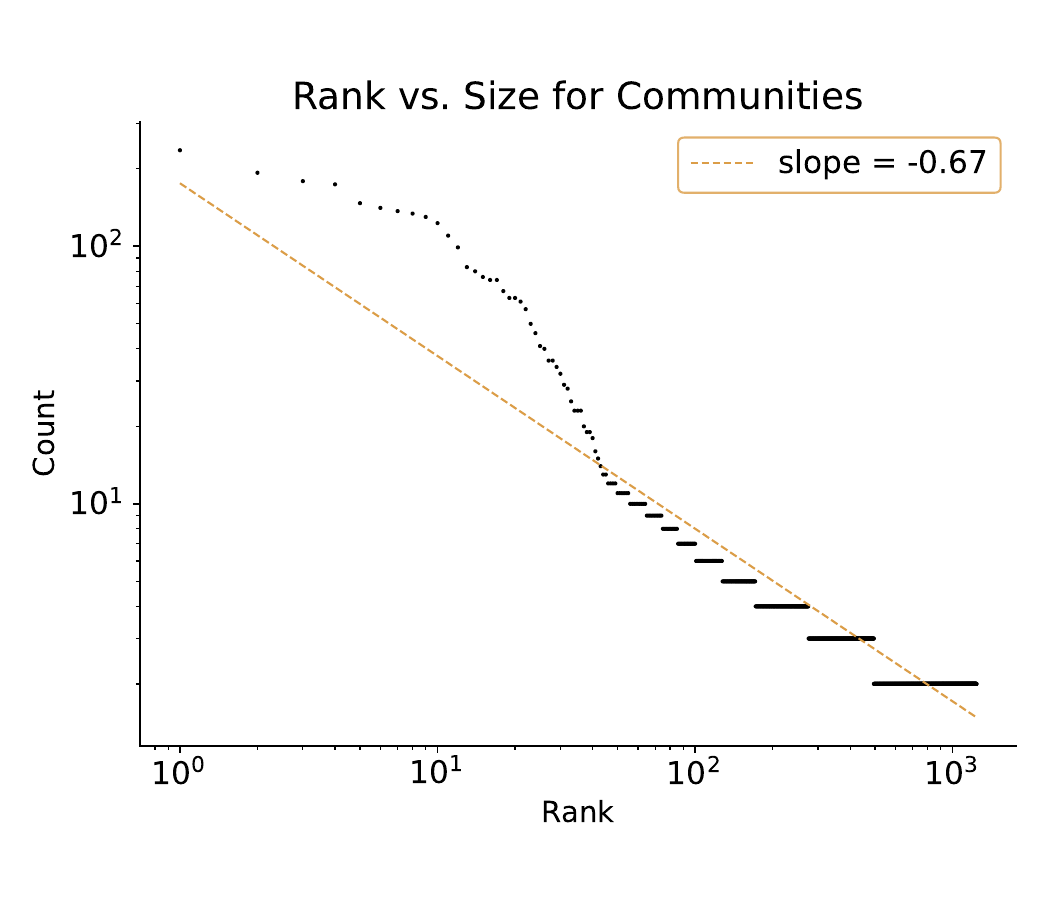}
\caption{The size of each community plotted against its rank, where rank=1 is the biggest community. We removed all communities containing only a single paper.}
\label{fig:zipf}
\end{figure}

\subsection{Communities}

For the community labels, we created a network of papers with each edge representing an author; two papers are connected if they share an author. To avoid small-world problems from the small number of papers with many authors, the weight of the edge between two papers $P_1$ and $P_2$ was defined as

$$W(P_1, P_2) = \frac{1}{len(P_1.\text{authors}) \times len(P_2.\text{authors})}$$

We ran Louvain community detection~\cite{blondel2008fast} to create the Community labels. Our community labels represent a community of papers linked by common authors, not a more traditional (and intuitive) community of authors linked by common papers. We are less interested in the institutional dynamics of misinformation scholars than the history of a concept. We used Louvain because our network has many nodes (13339 papers); the number of communities is not known prior to detection, and our analysis is only on the largest communities, meaning that the danger of  missing small communities with Louvain is a worthwhile trade-off for its speed and dynamically-chosen number of communities.

Our algorithm labeled 8033 distinct communities. Figure~\ref{fig:zipf} shows the distribution of sizes of communities. 6797 are communities of just one paper, and the average number of papers in a community is 1.66. Many of these one-paper communities are the result of authors using the term misinformation in passing. For example, \citet{konesPreventionFantasyFuture2011}'s ``Is Prevention a Fantasy, or the Future of Medicine? {{A}} Panoramic View of Recent Data, Status, and Direction in Cardiovascular Prevention'' contains the sentence ``Health, nutrition, and exercise illiteracy is prevalent, while misinformation and unrealistic expectations are the norm.'' In other words, these are not misinformation papers, but are instead using the term in some of the many ways discussed in Section~\ref{intro}.

We looked at the two biggest paper-communities in every year, as well as all the biggest communities in 2023, and analyzed the citation flows between communities, which we discuss in Section \ref{results}.

\section{Results}
\label{results}

\begin{figure*}[t]
\centering
\includegraphics[width=\textwidth]{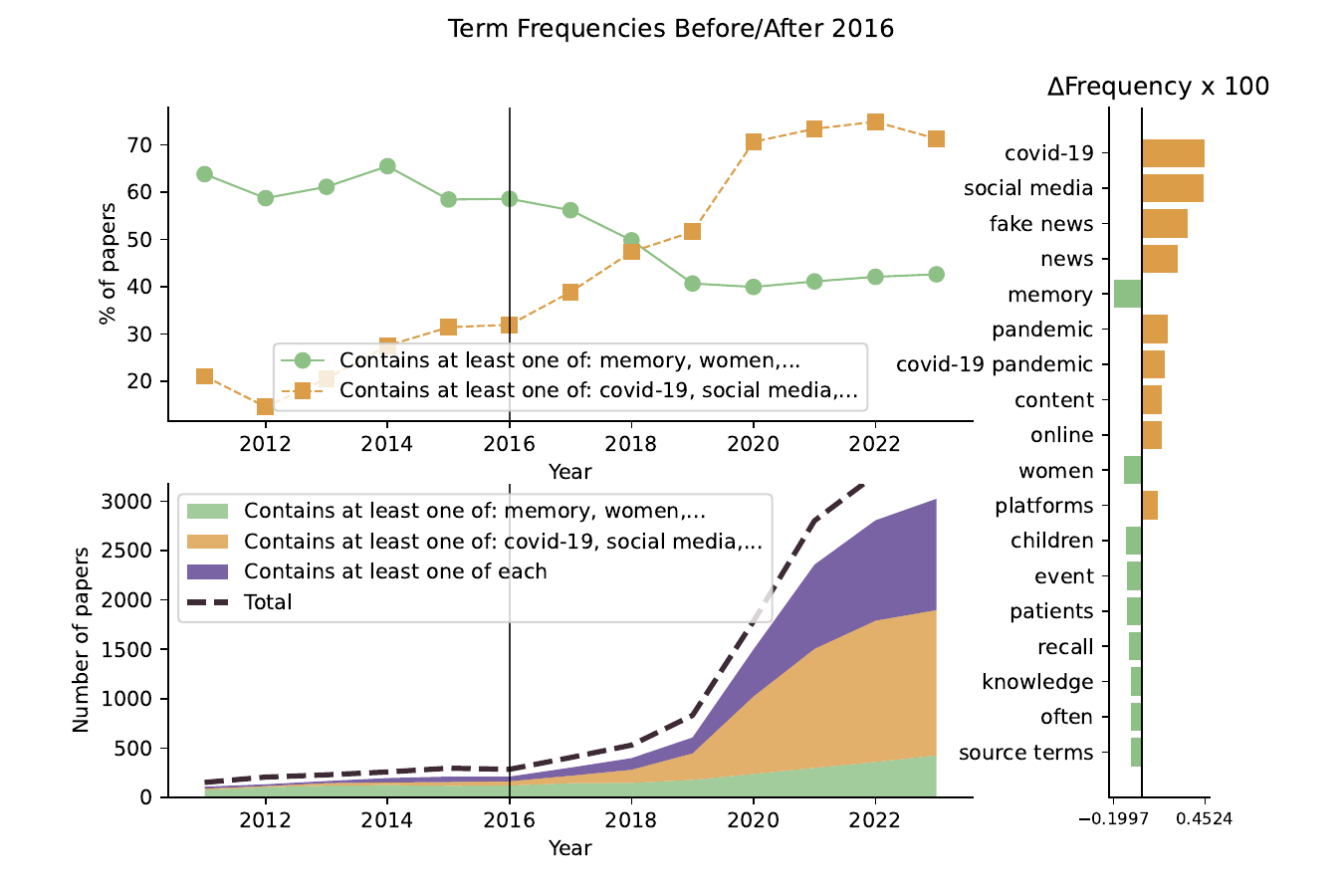}
\caption{Words with largest absolute frequency shifts before and after 2016 are on the right. We tagged all papers as having one from either or one of each. The top-left graph shows papers that have at least one word from that set of words as an annual percentage (some papers use both, so percentages do not add up to 100\%); the bottom one is the raw counts of all papers in the data.}
\label{fig:tf}
\end{figure*}

After 2016, there is an explosion of the use of the word ``misinformation'' in academic literature. In 2011, the first year in our dataset, there were only 118 papers containing the term anywhere in their metadata. In 2023, there were 3380, or 28 times as many, and the post-2016 corpus contains 20 times as many papers as the pre-2016 one.

Looking at terms with the largest absolute frequency shift (Figure \ref{fig:tf}), the post-2016 words are those that we in 2026 expect in misinformation literature, like ``social media,'' ``pandemic,'' ``fake news,'' etc. As discussed in our introduction, these findings on the post-2016 explosion, as well as the topics of study (e.g., social media and the pandemic), are similar to those found in literature reviews~\cite{broda_misinformation_2024}, histories of misinformation~\cite{donovan_express_2025}, and academic summaries~\cite{posetti_short_2018} of the misinformation literature.

To our knowledge, however, the pre-2016 terms remain largely unexplained in misinformation histories and literature reviews. For example, in their literature review, \citet{broda_misinformation_2024} used a similar date range in their analysis as our own (2010-2021). Their dataset comes from Google Scholar, and they used keywords ``misinformation,'' ``disinformation,'' and ``fake news.'' They too find a similar explosion in the literature, and on a similar timeline, but their summary does not account for words like ``women,'' ``children,'' ``memory,'' and ``recall,'' though there are hints of the same phenomenon. For example, when they break down the disciplines in which misinformation research is published, they find that the top 5 are Communication, Computer Science, Psychology, Political Science, and Economics. They note that the research in Psychology journals tends to focus on experiments, and that these experiments show broad themes. We discuss this in Section \ref{discussion}.

%This discrepancy seems to be, at least in part, a dicrepancy in data collection. They tried to find papers that had ``[a]t least one of the keywords ... present in title, keywords or abstract,'' but their Figure 1 shows virtually no papers until 2014. Even after, the number seems to be, at most, a couple dozen per year until past 2016. They also explain their difficulty in navigating Google Scholar's opaque algorithm, an increasingly common problem in computational studies~\cite{poudelCuratedRealitiesIntersection2025}. As discussed in Section~\ref{limitations}, this paper too suffers from a similar limitation, though from a different data source.

\begin{figure*}[t]
\centering
    \includegraphics[width=\textwidth]{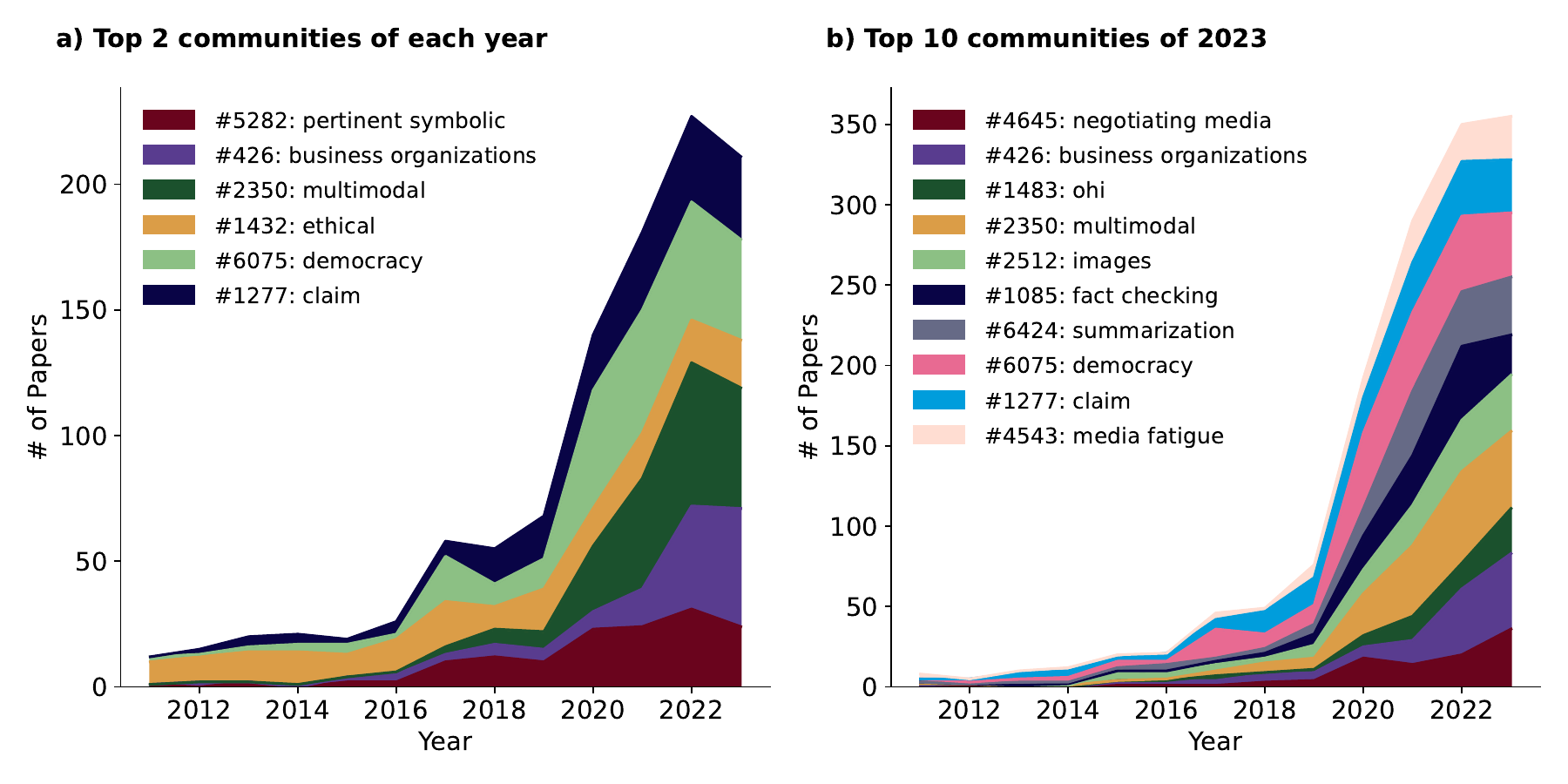}
    \caption{In both graphs, each community is labeled with its top TF-IDF term. In (a), we count papers from the two biggest communities of each year in the dataset throughout its duration (Note that the same community often appears in the top 2 for many years, which is why there are fewer than might be expected). In (b), we instead count papers in the top 10 paper-communities as of 2023. Most histories of misinformation view the field as in (b), but our paper is about the interpretation in (a). In (b), ``OHI'' stands for ``Online Health Information''.}
    \label{fig:communities}
\end{figure*}

Most summaries of the field tell its history as shown in the second half of Figure~\ref{fig:communities}: They look at the most prominent strains of research in the 2020s, then trace them back to their origin. Their analyses then flow from there. As a result, some histories explicitly define their inquiry to start with social media, defining a concept like ``misinformation-at-scale''~\cite{donovan_express_2025}.

\begin{table}[t]
\footnotesize
\begin{tabular}{p{0.2\columnwidth} p{0.3\columnwidth} p{0.5\columnwidth}}
\hline
Community & Top Unigrams & Top Bigrams 
\\
\hline
426 & governments \newline pandemic \newline trust \newline crises \newline business \newline unbuilding \newline organizations & business organizations \newline media governments \newline news media \newline pandemic research \newline health crises \newline arise 2024 \newline building unbuilding 
\\
\hline
1277 & claim \newline task \newline claims \newline false \newline media \newline participation \newline social & claim task \newline social media \newline spread false \newline fake news \newline 2023 forum \newline 2023 primary \newline 2023 uncovering \\
\hline
1432 & ethical \newline debriefing \newline researchers \newline research \newline dialogue \newline deception \newline insurmountable & ethical practices \newline ethical research \newline adopting reporting \newline balance deception \newline benefits individual \newline challenges insurmountable \newline conducting ethical 
\\
\hline
2350 & multimodal \newline dis \newline disciplines \newline review \newline computer \newline science \newline analysis & multimodal analysis \newline multimodal dis \newline computer science \newline 2020 start \newline analysis dis \newline attention academics \newline beyond words 
\\
\hline
5282 & symbolic \newline pertinent \newline content \newline language \newline linguistic \newline representative \newline detection & pertinent symbolic \newline misleading content \newline content based \newline real world \newline accuracy also \newline additional training \newline aggregating representative 
\\
\hline
6075 & democracy \newline knowledge \newline shared \newline discredited \newline frontier \newline countermeasures \newline undermining & attacks involves \newline campaigns undermining \newline citizens example \newline contribute countermeasures \newline countermeasures 2025 \newline democracy democracy \newline democracy relies \\
\hline
\end{tabular}

\caption{Top TF-IDF terms for select communities. Note how the Loftus community's terms differ from those of the other communities, the rest of which seem more familiar to the post-2016 misinformation paradigm} 
\label{tab:tfidf}
\end{table}

\begin{figure*}[t]
\centering
\includegraphics[width=\linewidth]{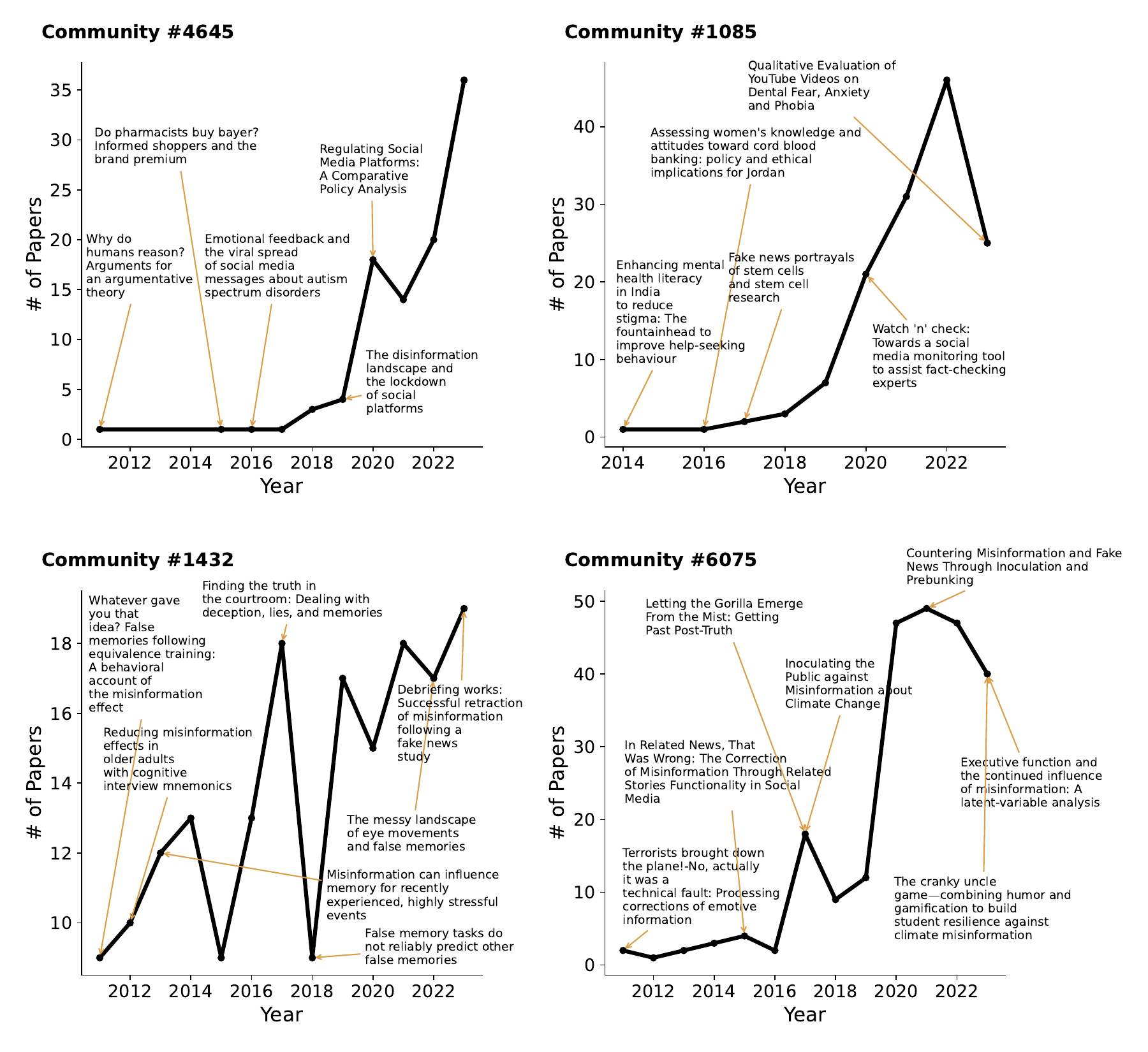}
\caption{Four communities' papers, including the Loftus community (\#1432), plotted throughout the duration of our dataset. Note the diverse starting points and the gradual evolution towards post-2016 paradigmatic misinformation research.}
\label{fig:examples}
\end{figure*}

This commonly-employed nihilogony ignores the several strands of research the post-2016 paradigm ties together. If, instead of looking at the biggest communities of 2023 back through time, we take the two biggest ones of every year and plot those through the duration of our dataset, we find that, of the pre-2016 strands, community \#1432 (which we will refer to from now on as the ``Loftus community,'' after its most cited author) is the biggest and most well-established. This is the community associated with the aforementioned misinformation effect (see Section~\ref{intro}). It maintains a steady size across these years, even as topics like ``fake news'' and ``covid'' rose in popularity and prominence around it. These papers primarily work in psychology, with a particular focus on the effects of misinformation on memory.

Table ~\ref{tab:tfidf} contains the top TF-IDF scores for select paper communities. Those top terms (``ethical,'' ``debriefing,'' ...) in the Loftus community's papers may, at first, seem unrelated to those of the pre-2016 research found in Figure \ref{fig:tf}. To make the connection, consider this abstract from our database, from~\citet{otgaar_protecting_2020}, in a paper titled ``Protecting Against Misinformation: Examining the Effect of Empirically Based Investigative Interviewing on Misinformation Reporting,'' published in \textit{Journal for Police and Criminal Psychology}. Our community detection put this paper in the Loftus community.

\begin{quote}
    Children who are involved in legal cases are often interviewed about events they witnessed or that might have happened to them. [...] We found that children’s recall during the NICHD interview protected children against the incorporation of misinformation in their accounts of the event [...] The current experiment suggests that evidence-based investigative interviewing can inoculate children’s memory against the corrupting impact of misinformation.
\end{quote}

In our database, the Loftus community is the only well-established pre-2016-paradigm strain of misinformation research, one that explains the pre-2016 frequency shifts in terms. We also find that it is being incorporated into the post-2016 paradigm in two senses: Its research is increasingly similar to that of other misinformation paradigm papers, and its concepts and framings can be seen influencing the other paper-communities. 

Figure ~\ref{fig:examples} shows four communities of papers, each of which, save for the Loftus community, was very small pre-2016. These communities ultimately converge on recognizable misinformation research, though each brings their own concepts and framings. In \#4645, we can see research that began with an interest in cognition in its most general terms, then in the impact of branding on pharmaceuticals. From there, it moves into how social media users talk about autism, until containing the misinformation research recognizable as the post-2016 paradigm. Similarly, \#1085 moves from pre-2016 questions of different kinds of health literacy in the public (e.g., mental health or cord blood banking), to more paradigmatic questions of social media monitoring and fact checking.

\begin{figure}[h]
\centering
\includegraphics[width=1\linewidth]{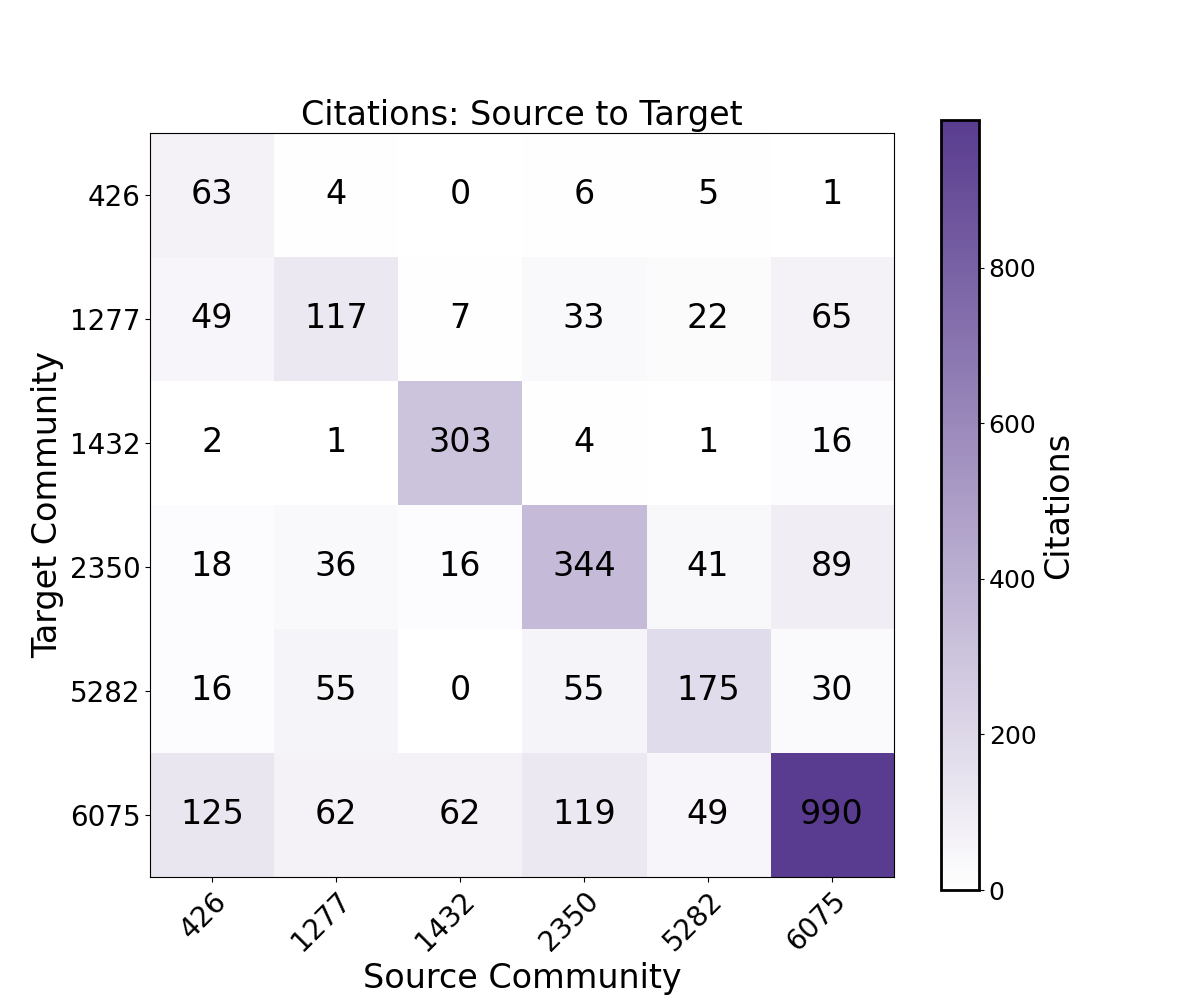}
\caption{A citation flows from a ``source'' paper to a ``target'' paper if the ``source'' paper cites the ``target'' paper in its references.}
\label{fig:citations}
\end{figure}

\begin{table*}[t]
  \begin{tabular}{p{.40\textwidth}p{0.02\textwidth}p{.40\textwidth}}
%% \begin{tabular}{P{.40\textwidth}P{.40\textwidth}}
\toprule
Source Paper & & Target Paper \\
\midrule
Misinformation and Worldviews in the Post-Truth Information Age: Commentary on Lewandowsky, Ecker, and Cook & & Believing is Seeing: Biased Viewing of Body-Worn Camera Footage \\
\midrule
Countering Misinformation and Fake News Through Inoculation and Prebunking & & Aging in an Era of Fake News \\
\midrule
Countering Misinformation and Fake News Through Inoculation and Prebunking & & False Memories for Fake News During Ireland’s Abortion Referendum \\
\midrule
Psychological Inoculation against Misinformation: Current Evidence and Future Directions & & Nevertheless, partisanship persisted: fake news warnings help briefly, but bias returns with time \\
\midrule
Misinformation: susceptibility, spread, and interventions to immunize the public & & Aging in an Era of Fake News \\
\midrule
Examining the replicability of backfire effects after standalone corrections & & Vaccination against misinformation: The inoculation technique reduces the continued influence effect \\
\midrule
Tackling misinformation: What researchers could do with social media data & & Aging in an Era of Fake News \\
\midrule
How Official Social Media Affected the Infodemic among Adults during the First Wave of COVID-19 in China & & Pandemics and infodemics: Research on the effects of misinformation on memory \\
\midrule
AI Art and Misinformation: Approaches and Strategies for Media Literacy and Fact Checking & & Vaccination against misinformation: The inoculation technique reduces the continued influence effect \\
\bottomrule
\end{tabular}

\caption{
An example of citations flowing from outside the Loftus community papers ``sources'' to inside the Loftus community ``targets.''  The former's titles tend towards language and framings more familiar to the post-2016 misinformation paradigm, and the latter's tend towards framing from the perspective of individual cognition. However, the two show clear overlapping interests and mutual engagement.}
\label{table:citations}

\end{table*}

Figure~\ref{fig:citations} shows a small amount of citations flowing into the Loftus community from other communities. The small number is, in part, an artifact of our data, because the Loftus community is a smaller community than the others in that figure, but also because our data collects only papers with ``misinformation'' in metadata, and the influence often flows through other concepts. 

Table~\ref{table:citations} contains a selection of papers outside the Loftus community citing papers within it. Loftus commmunity papers tend to concern themselves primarily with the individual, cognitive effects of having heard misinformation, whereas post-2016 misinformation researchers might more typically focus on how it circulates in social media. These are complimentary views, often examining identical phenomena from slightly different perspectives. For example, \citet{brashierAgingEraFake2020}'s ``Aging in an Era of Fake News'' examines the susceptibility of older adults to misinformation on social media.

Sometimes, the different provenances of ideas cause friction. Consider the response from \citet{hymanMisinformationWorldviewsPosttruth2017} to \citet{lewandowskyMisinformationUnderstandingCoping2017}. From the former:

\begin{quote} 
    Based on combating misinformation, Lewandowsky et al. made narrow recommendations for future cognitive research, including the generalization of inoculation techniques [...] Unfortunately, while Lewandowsky et al. acknowledged that world views matter, they argued against addressing worldviews. They stated that for corrections to be effective “they must not directly challenge people’s worldview” [...] [They] were concerned with the ethical implications that come with trying to modify individuals’ world views. Therefore, Lewandowsky et al. argued that “worldview ought to be decoupled from the propensity to accept misinformation and resist its correction”.
  \end{quote}

This debate might be seen as an attempt to resolve a tension between two slightly divergent understandings of the target phenomenon of study. The post-2016 misinformation paradigm is intertwined with electoral politics and social media. In that context, changing people's worldviews, or perhaps stopping them from being changed by unscrupulous actors, is core to the project. We see that mirrored in \citet{hymanMisinformationWorldviewsPosttruth2017}'s emphasis on the worldview as one of the (if not the most) salient question in studying misinformation. Democracy cannot function without deliberation, the point of which is to understand and change people's minds. As they put it, ``we believe the problem to be addressed is more substantive than people simply adopting some of the misinformation available in our post-truth world.''

The Loftus community's work, by contrast, emphasizes eye witness accounts in therapy sessions and courtrooms. If eye witnesses to crimes exhibit that same open-mindedness about the events that transpired, their testimonies are useless. In that vein, \citet{lewandowskyMisinformationUnderstandingCoping2017} opens with a reference to misinformation effect research, in which "changing a memory of a stop sign to a yield sign was a remarkable demonstration of the power of misinformation."

Despite some inevitable friction, the varying conceptions of misinformation are, by and large, complimentary, allowing for bridging concepts. They are often, for example (and as \citet{hymanMisinformationWorldviewsPosttruth2017} note), interested in the efficacy of different techniques for stopping people from believing misinformation, or of ``prebunking''/``inoculating against misinformation'' (as the examples in Table~\ref{table:citations} call it). In papers like \citet{gradyNeverthelessPartisanshipPersisted2021}'s ``Nevertheless, Partisanship Persisted: Fake News Warnings Help Briefly, but Bias Returns with Time'' and \citet{murphyFalseMemoriesFake2019}'s ``False Memories for Fake News During Ireland's Abortion Referendum'' (both co-authored by Loftus), we see research that comfortably fuses the concerns of post-2016 misinformation paradigm of fake news and bias with the older focus on the cognitive effects of the misinformation. 

Post-2016 misinformation research has studied the effects of fake news on memory or belief formation enough that there is at least one literature review on that specific subtopic~\cite{faedda_fake_2024}.  

We can also detect signs of these more diffuse indicators of influence in our database, where we find 157 papers matching the regex \texttt{.*[ -]formation.*}, the majority of which are discussing ``belief formation,'' ``opinion formation,'' etc. Of these, only ten are from the Loftus community, suggesting conceptual ties that extend beyond what the citation flows might show.

We can see these same patterns in Loftus's own work. In a paper titled ``From Primary to Presidency: Fake News, False Memory, and Changing Attitudes in the 2016 Election,'' Loftus is a co-author in an example of the older paradigm intertwined with the new one~\cite{grady_primary_2023}. From their abstract:

\begin{quote}
    [...] This study followed 602 United States citizens, recruited from Amazon Mechanical Turk, at three points throughout the 2016 presidential election investigating how attitudes and preferences changed over time and how people remembered their past feelings. Across political parties, people’s memory for their past attitudes was strongly influenced by their present attitudes; more specifically, those who had changed their opinion of a candidate remembered their past attitudes as being more like their current attitudes than they actually were. Participants were also susceptible to remembering false news events about both presidential candidates. However, they were largely unaware of their memory biases...
\end{quote}

This is a classic Loftus community experimental setup, but merging with the new language of fake news and data-gathering through clickwork services, a common practice in computer science research.

\section{Discussion}
\label{discussion}

With one major caveat, which is the principle subject of this paper, the results discussed in Section~\ref{results} are ultimately similar to those of from \citet{broda_misinformation_2024}:

\begin{quote}
In the case of misinformation, disinformation, and fake news, it also appears as if different strands of literature have developed quite independently of each other. The result is a rather scattered body of work.
\end{quote}

We found, in agreement with the above, that these big communities of papers are mostly self-citing, and, if they existed at all before 2016, it was in some primordial state. This also agrees with many other reviews and histories, many of which explicitly link the concept to social media, even when they gesture broadly to vague historical context~\cite{donovan_express_2025,perez-escolar_systematic_2023,miro-llinares_misinformation_2023, faedda_fake_2024}. These short passages tend to take for granted the concept of misinformation itself, operating exclusively within the paradigm. \citet{faedda_fake_2024} give us a typical example (reproduced with citations removed):

\begin{quote}
    Although misinformation (i.e., information that is erroneous or misleading) can circulate exceedingly fast due to advances in social technologies and large-scale information cascades, the roots of fake news go back to the days before the Printing Revolution, when word-of-mouth was the primary method of news transmission. 
\end{quote}

In the Loftus community, however, we find a caveat that, to our knowledge, goes unacknowledged in other reviews and histories of misinformation. This is a strain of research that much predates 2016, one that was established prior to the beginning of our dataset (2011) and had played a prominent role in public life. 

In Section~\ref{concept}, we will consider the one through line that unites this otherwise-disparate set of research, the concept of misinformation itself, with the accompanying way it models the world. In Section~\ref{history}, we will examine the historical context from which the concept emerges.

\subsection{Misinformation}
\label{concept}

As its core components (whose provenance we discuss in Section~\ref{history}), the misinformation paradigm considers individuals making, reading, and sharing false statements, usually on social media, producing research that often warns of the harm to society resulting from the sum of many such events. This approach is in no way self-evident. Given that, as discussed throughout this paper, most authors agree that the advent of social media is central to this line of inquiry, it could have been that the field coalesced around the political economy of social media, as \citet{chomsky1988manufacturing} did for mass media in 1988. 

Alternatively, misinformation scholars might have followed \citet{williamsTelevisionTechnologyCultural2004}'s example. In his 1973 analysis of television, he brings together the technology of broadcasting and the cultural form of television to describe the ``flow'' state of watching television, a state that allows for a simultaneously communal yet private experience. There is a parallel here to ``doomscrolling''~\cite{priceDoomscrollingCOVID19Negative2022}, a popular concept that social media users deploy to explain their own behavior, yet we rarely find discussion in this vein.

In general, misinformation research tends to see through the medium, emphasizing the directness of interactions. \citet{cinelliCOVID19SocialMedia2020} provide us with a typical example:

\begin{quote}
Social media platforms such as YouTube and Twitter provide direct access to an unprecedented amount of content and may amplify rumors and questionable information
\end{quote}

This is a major departure from \citet{mcluhan_understanding_media_1964}'s ``the medium is the message.'' Similarly, whereas \citet{williamsTelevisionTechnologyCultural2004} took great pains to thoroughly describe and theorize the many design decisions that went into television, we have yet to read a comparable account to, for example, the Twitter user experience in misinformation literature, though other researchers have brought this line of work to the study of social media~\cite{scharlachVALUEAFFORDANCESSOCIAL2023}. When misinformation researchers do talk about social media as active agents, not just as a passive conduit for pseudo-direct interactions, they often discuss the platforms as potential collaborators in stopping misinformation, like, for example, developing tools to detect it~\cite{broda_misinformation_2024}. There is rarely mention of the legions of designers, engineers, product managers, etc. that designed social media to work the way it does, or the reasons that they did so. Upon encountering social media companies solely through misinformation literature, one might think that their role is limited to passively waiting for academics to find a technical solution to misinformation, probably some sort of detector, but perhaps also some kind of media literacy training program.

Despite its emphasis on social media, the post-2016 misinformation paradigm looks less like media theory and more like the Loftus community's psychological research. There are modern scholars that continue the tradition of media research on social media~\cite{eichhornContent2022, scharlachVALUEAFFORDANCESSOCIAL2023}, but one rarely finds references to it. Instead, misinformation research focuses on the content, sources, and consumers of misinformation, not its context, cultural form, or structural factors. It is probably no coincidence that there is a parallel here with social media database design, whose schema are also comprised of those same objects, a resonance we hope to explore in future work. In short, when compared to misinformation effect research, there is a change in emphasis from cognition to the misinformation itself, along with an increase in scale, but the core model is recognizable. 

This shift in emphasis, however, brings difficulties. On a theoretical level, the core model of misinformation effect research was taken from the social processes of the Satanic panic. Their research attempts to isolate and understand the false memories that can be implanted by leading questions from police investigators or recovered memory therapists. In its new context, the model loses that grounding.

On a practical level, researchers within the misinformation paradigm commit themselves to evaluating individual pieces of (mis)information, a theoretical commitment that becomes difficult to reconcile with the scale of social media. As a result, researchers employ simplifying assumptions. For example, they often define misinformation as a social media post that includes links to known unreliable sources~\cite{flaminoPoliticalPolarizationNews2023, bovetInfluenceFakeNews2019, hanMeasuringCoevolutionOnline2025, guessExposureUntrustworthyWebsites2020}, and often with a note on how this is a common choice in the literature, e.g., ``Following
previous research, we used a domain-level approach to measurement''~\cite{guessExposureUntrustworthyWebsites2020}. 

Some researchers themselves acknowledge the problem here. \citet{hanMeasuringCoevolutionOnline2025}, for example, both deploy this simplification and acknowledge its shortcomings (emphasis added):

\begin{quote}
    We annotate the scientific quality of retweets based on the web domains they reference —a measure that is also \textbf{commonly used in other research}, rather than evaluating the content of the individual articles themselves. We recognise that the latter allows the classification of a broader range of content, including posts without web links, and can be more accurate in some cases. Although incorporating the content-base analysis into our investigation would be valuable, it would be more resource-intensive, potentially posing challenges for timely monitoring implementation.

\end{quote}

Defining misinformation as something that contains a link to an unreliable site is an oversimplification in two complimentary ways. First, it ignores anything that is not a direct link to a source of misinformation. This misses not just many instances of misinformation-sharing, as the above quote notes, but also important dynamics in how peer-to-peer information sharing (i.e., the core functionality of social media) actually functions. Attempts to quantify how much of media consumers' information comes directly from fake news sources argue that they are minuscule when compared to other factors~\cite{Allen2020FakeNews}, a result that casts doubt on results relying on these simplifications.

Second, this model assumes that everything unreliable sources say is untrue. This contradicts the documented best practices of those professionally employed to shape public opinion. Practitioners of public relations are trained to avoid lying~\cite{tremblay_pepin_intrigues_2025}. This has been best practice in governments since at least the 1940s. As the Supreme Headquarters Allied Expeditionary Force of World War II put it~\cite{ellul1973propaganda}:

\begin{quote}
    When there is no compelling reason to suppress a fact, tell it. [...] Aside from considerations of military security, the only reason to suppress a piece of news is if it is unbelievable [...] When the listener catches you in a lie, your power diminishes [...] For this reason, never tell a lie which can be discovered.
\end{quote}

If researchers wish to study how it is that the public can be deceived, or at least dishonestly influenced, their simplifying assumptions probably ought not dismiss the actual practices of those who do public relations or military propaganda, who operate in the chasm between true and false, rather than at its poles. Similarly, the more likely that a piece of misinformation is to be believed, the more difficult it necessarily is to detect as misinformation. People do not like being fooled, purposefully or otherwise. If untenable theoretical commitments force misinformation scholars to make these simplifying assumptions, then they risk focusing their inquiry precisely on the least salient part of the phenomenon.

To be clear, our argument is not that the post-2016 misinformation paradigm inherits exclusively from the Loftus community. In future work, we hope to discuss other factors that shaped the paradigm. Instead, the point here is that, were the provenance of its framework different, it is difficult to imagine that misinformation researchers would have made these same theoretical commitments. As discussed further in Section~\ref{conclusion} (though it bears repeating), the purpose of this paper is not to discredit science as such, especially under the circumstances, nor is it to argue that science is powerless to investigate these processes. Instead, we hope that, by bringing these histories to light, both scholars and the public can better navigate the challenges that misinformation researchers hope to address~\cite{porter_how_2009}.

\subsection{History}
\label{history}

To better understand these theoretical commitments and their provenance, we must briefly summarize the Satanic panic. Though many of the cultural threads that formed the Satanic panic already existed~\cite{emerson2024unmask}, its mainstream history begins in 1980, when Michelle Smith co-authored \textit{Michelle Remembers} with her psychiatrist and soon-to-be-husband, Lawrence Pazder. In it, she recounts her gruesome experience being tortured at the hands of a Satanic cult as a very young girl, memories that, until recently, she had repressed. Pazder had used modern psychiatric techniques to help her recover these lost memories. These stories turned out to be grotesque fabrications, but the book, along with its credulous media coverage, triggered what became known as the Satanic panic, in which innocent daycare workers, teenagers who liked hard rock, and others were convicted of ritual child abuse~\cite{hearst_qanon_2022,shewan_conviction_2015}.

A full account of the panic is out of scope for this paper. However, in order to illustrate the continuity of both misinformation scholarship and the cultural phenomena that motivated it, we will briefly discuss the McMartin Preschool trials, one of its most infamous incidents. In one of the longest and most expensive trials in US history, the school's owners and staff were accused of ritually abusing of hundreds of children as part of their Satanic worship~\cite{Reinhold1990LongestTrial, linderMcMartinPreschoolAbuse2007}. The police investigation began with the report of one parent, after which they sent a letter to the 200 families at the school asking for information. The letter was shockingly explicit, saying that their children might have been victims of sexual abuse, outlining the potential abuse in some detail, adding that they may have been photographed, and asking parents for any additional information~\cite{Reinhold1990LongestTrial}. This set off a chain of events that elicited numerous false confessions from children which began with the parents and ended with the police systematically coercing confessions from many of the supposed victims~\cite{Schreiber2006Suggestive}. 

Rather than show restraint, the media, like the police, amplified the salacious allegations, both in the specific case of the McMartin trial~\cite{shaw2026}, as well as throughout the panic more generally~\cite{SuburbanUncanny2019}. In another striking parallel to modern conspiricism, there were allegations that the abuse took place in hidden tunnels underneath the school~\cite{WildClaimsMass2024}, mirroring the modern allegations about the similarly non-existent basement of Comet Ping Pong~\cite{bentonEverythingOldAgain2025}.

 As discussed in Section~\ref{intro}, Elizabeth Loftus, the Loftus community's most cited author, studies the misinformation effect. Much of this research was a response to the Satanic panic, or, as Loftus put it in 1993,~\cite{loftus_reality_1993} ``a rise in reported memories of childhood sexual abuse that were allegedly repressed for many years.'' Her work explicitly references the Satanic panic, and considers not just the cognitive phenomenon of the misinformation effect, but, from the abstract from that same 1993 paper...

\begin{quote}
    (a) How common is it for memories of child abuse to be repressed? (b) How are jurors and judges likely to react to these repressed memory claims? (c) When the memories surface, what are they like? and (d) How authentic are the memories? 
  \end{quote}

As we have seen, Loftus was a prominent skeptic of the recovered-memory therapy behind the Satanic panic of the 1980s and 1990s, and has continued to publish on memory, narrative, and misinformation. More recently, as discussed in Section~\ref{results},
she has published work considering ``fake news'' as potentially influencing false memories, e.g., ``Misinformation–past, present, and future''~\cite{loftus_misinformation_2024}.

It would be an oversimplification to say she was inspired to do this research because of the Satanic panic. As previously discussed, her work on these issues began in the early 1970s, predating it. Likewise, as Figures~\ref{fig:communities} and \ref{fig:examples} show, the post-2016 misinformation paradigm does not come from nothing. Authors like~\citet{nguyen_sources_2012} were already investigating sources of misinformation on social media. In both cases, social changes met with researchers' prior interests, and the two mutually affected each other, creating and growing a paradigm of public interest. 

The post-2016 misinformation paradigm comes at a time when people who believe that the world is run by a cabal of child-torturing pedophiles (i.e., Qanon) are a powerful political movement, forming a core part of the current US administration's coalition, and whose influence has penetrated deep into public life~\cite{feeld_qanon_2025}. Because Qanon is a social media phenomenon, and because the misinformation paradigm was, in a measurable sense, born in 2016 (see Figures~\ref{fig:tf}~and~\ref{fig:communities}), it makes sense to begin its history there. To view the field this way, however, is to deprive its practitioners of valuable historical context, one with striking parallels to 2026, a rich vein from which we can mine insights about the role of science in public life, and even about its relationship to new forms of mass media. More specifically, we note the following five parallels. 

First, then as now, technical advances (in recovered memory techniques and telecommunications) were linked to misinformation crises. Second, both crises spread through recent innovations in mass media~\cite{hearst_qanon_2022}. Third, misinformation researchers mobilized science to respond to these moments of acute conflict over epistemic authority~\cite{Zagzebski2012}. Post-2016 misinformation research often discusses this by noting the so-called ``post-truth'' era, a term that appears in abstracts 133 times in our database.
  
Fourth, researchers sought to understand and combat the harm, often participating both in public debate and in the very social processes they study. Misinformation literature in computer science journals, for example, often proposes ways to automatically detect misinformation on social media~\cite{broda_misinformation_2024} while also seeing misinformation as a social media phenomenon. Likewise, Loftus has participated in the legal system, often as an expert witness for the defense in high profile trials~\cite{shewan_conviction_2015}. In these struggles, scientists participate in boundary work~\cite{gierynCulturalBoundariesScience1999}, attempting to simultaneously wield the epistemic authority of science while defining its borders to exclude those whose input they (often rightly) deem beneath it, such as recovered-memory practitioners, conspiracists, and vaccine skeptics. 

Fifth, in both instances, there was widespread social concern that powerful malevolent forces were murdering children~\cite{hearst_qanon_2022,shewan_conviction_2015}.

\subsection{Conclusion}
\label{conclusion}

As the TF-IDF terms in Table~\ref{tab:tfidf} show, many of the papers in our database warn of democracy under threat, or of distrust in science and/or institutions. Some authors argue that misinformation and distrust in institutions are linked~\cite{akyuz_impact_2021}, or that misinformation causes distrust in institutions~\cite{stetka_have_2025}. To our knowledge, it is much less common to question whether our institutions have earned the public trust, and, if not, how our institutions ought to behave in moments of contention for epistemic authority.

In the history of misinformation, we see the social role of science in both its successes and its failures. Scientific techniques gave us \textit{Michelle Remembers}, and skeptics in the scientific community took to testing whether these memory techniques were indeed reliable. Lest anyone think that this story is a simple one of the scientific method working as intended, one in which dispassionate scientists debunk a bad hypothesis, the way that science is often said to work, Loftus is a controversial figure. In her career as a public intellectual, she has, for example, provided expert testimony on the unreliability of memories for Harvey Weinstein's~\cite{press_harvey_2020}, Ted Bundy's~\cite{marsh_how_2012}, and Ghislaine Maxwell's~\cite{news_ghislaine_2021} defenses. 

Going beyond Loftus, though recovered memory therapies are now debunked, the ``Memory Wars'' are not definitively settled~\cite{otgaarReturnRepressedPersistent2019}, and, as the scientific debate goes on, innocent people's lives have been shattered by the ideas it generated~\cite{shewan_conviction_2015}. As \citet{porter_how_2009} puts it, ``The role of the sciences in regard to public issues of all kinds has never been more encompassing.'' As we have seen, juries depend on scientists to know if witnesses are reliable, and psychiatric theories are deployed on patients seeking help in their most vulnerable moments, sometimes to disastrous effect. This makes ``the divide between technical science and political opinion is highly unstable''~\cite{porter_how_2009}.

To that end, searching for the regex \texttt{.*[ -]detect.*} returns 1500 abstracts in our database, because misinformation detection is a popular idea in the scientific literature~\cite{broda_misinformation_2024}, one that would necessarily rely on social media companies for its implementation. This research places science in a technical role at the interstices of politically-charged social functions, one that once again has historical parallels, this time in Loftus's expert testimony in the courtroom. 

We mention Loftus's work on legal defenses not to criticize her (to the contrary, the primary author of this paper has found her work, on the whole, compelling), but to illustrate the complex role of the scientist in public life. Despite the recent success enjoyed by its opponents, to speak with the authority of science is still to wield considerable power. 

\subsection{Limitations}
\label{limitations}

With the exception of citations, all our data comes from a Scopus export. Much computational social science work relies on data derived from opaque, proprietary systems, and therefore risks systematic distortion~\cite{poudelCuratedRealitiesIntersection2025}. In our case, Scopus's search is opaque to us. Our arguments, however, rely on the existence of the data, especially the pre-2016 data, not necessarily its completeness. 

Our methodology also depends entirely on a single keyword. It could have been that researchers using the post-2016 misinformation paradigm simply happened upon the same term as those in the Loftus community with no conceptual ties. Conversely, researchers might, and in fact sometimes do, use a different term entirely while drawing heavily on the Loftus community's work, either directly or indirectly. We have argued only that there exists a lineage, showing ties that extend beyond citations, including through the concepts discussed in various papers. There may be other, important strands of pre-2016 research that we have not identified. Our primary point is to contextualize recent misinformation research by bringing attention to the influence of the misinformation effect.

Finally, our analysis restricted itself to the academic literature. As Figure~\ref{fig:twitter} shows, the term's rise in popularity is in no way exclusive to the academy. Just as science influences popular culture, so too are scientists participants in and influenced by, for example, the popular press or social media. Though they cannot be compared directly due to incompatible timescales, comparing Figure~\ref{fig:twitter} to Figure~\ref{fig:tf} suggests a process that goes beyond the academy. We hope in future work to further explore how and why the term came to define the paradigm, work that would require a more expansive dataset that goes beyond journal articles.

\subsection{Acknowledgments}
\label{acks}

The authors wish to express their appreciation to every student, faculty member, and staff member of the Vermont Complex Systems Institute. In particular, we express our appreciation to Jeremy Elliott for his involvement in this paper's precursor project, and to Sam Zhang, whose thorough feedback much shaped the paper's later drafts, much as his thoughtful approach to science continues to influence its first author. 

Thanks also to Nani Ruiz for many, many hours of conversation untangling (and re-tangling) the many concepts that plague us, relevant or otherwise, without which this paper would probably have not happened. 

Thanks to Ben Harris for many of the media theory recommendations that much influenced Section~\ref{concept}, chief among them ~\citet{williamsTelevisionTechnologyCultural2004}.

Thanks to Simon Tremblay-Pepin, whose influence on the first author extends well beyond the one citation herein and includes, among too much to list, providing a model of what scholarship could be in moments such as these.

The authors acknowledge financial support from MassMutual Center of Excellence in Complex Systems and Data Science, Google Open Source, and The National Science Foundation award \#2242829 (JWZ, DB, CMD, PSD) and \#2419830 (AJRI and CMD).

\bibliography{refs}

@article{broda_misinformation_2024,
	title = {Misinformation, {Disinformation}, and {Fake} {News}: {Lessons} from an {Interdisciplinary}, {Systematic} {Literature} {Review}},
	volume = {48},
	issn = {2380-8985},
	shorttitle = {Misinformation, {Disinformation}, and {Fake} {News}},
	url = {https://doi.org/10.1080/23808985.2024.2323736},
	doi = {10.1080/23808985.2024.2323736},
	number = {2},
	urldate = {2025-12-10},
	journal = {Annals of the International Communication Association},
	author = {Broda, Elena and Strömbäck, Jesper},
	month = jun,
	year = {2024},
	pages = {139--166},
}

@article{porter_how_2009,
	title = {How {Science} {Became} {Technical}},
	volume = {100},
	url = {https://doi.org/10.1086/599552},
	doi = {10.1086/599552},
	
	number = {2},
	journal = {Isis},
	author = {Porter, Theodore M.},
	month = jun,
	year = {2009},
	note = {Publisher: University of Chicago Press},
	pages = {292--309},
}

@article{zimmer_for_2025,
	chapter = {Science},
	title = {For {Real}, a {Natural} {History} of {Misinformation}},
	issn = {0362-4331},
	url = {https://www.nytimes.com/2025/12/09/science/evolution-fish-misinformation.html},
	urldate = {2025-12-10},
	journal = {The New York Times},
	author = {Zimmer, Carl},
	month = dec,
	year = {2025},
}

@article{perez-escolar_systematic_2023,
	title = {A {Systematic} {Literature} {Review} of the {Phenomenon} of {Disinformation} and {Misinformation}},
	volume = {11},
	issn = {2183-2439},
	url = {https://www.cogitatiopress.com/mediaandcommunication/article/view/6453},
	doi = {10.17645/mac.v11i2.6453},
	
	number = {2},
	urldate = {2025-12-10},
	journal = {Media and Communication},
	author = {Pérez-Escolar, Marta and Lilleker, Darren and Tapia-Frade, Alejandro},
	month = apr,
	year = {2023},
	pages = {76--87},
}

@article{miro-llinares_misinformation_2023,
	title = {Misinformation about fake news: {A} systematic critical review of empirical studies on the phenomenon and its status as a ‘threat’},
	volume = {20},
	issn = {1477-3708},
	shorttitle = {Misinformation about fake news},
	url = {https://doi.org/10.1177/1477370821994059},
	doi = {10.1177/1477370821994059},
	
	number = {1},
	urldate = {2025-12-10},
	journal = {European Journal of Criminology},
	author = {Miró-Llinares, Fernando and Aguerri, Jesús C.},
	month = jan,
	year = {2023},
	note = {Publisher: SAGE Publications},
	pages = {356--374},
}

@article{loftus_reality_1993,
	title = {The reality of repressed memories},
	volume = {48},
	issn = {1935-990X},
	doi = {10.1037/0003-066X.48.5.518},
	number = {5},
	journal = {American Psychologist},
	author = {Loftus, Elizabeth F.},
	year = {1993},
	note = {Place: US
Publisher: American Psychological Association},
	pages = {518--537},
}

@misc{shewan_conviction_2015,
	title = {Conviction of {Things} {Not} {Seen}: {The} {Uniquely} {American} {Myth} of {Satanic} {Cults}},
	shorttitle = {Conviction of {Things} {Not} {Seen}},
	url = {https://psmag.com/social-justice/make-a-cross-with-your-fingers-its-the-satanic-panic/},
	urldate = {2025-12-10},
	journal = {Pacific Standard},
	author = {Shewan, Dan},
	month = sep,
	year = {2015},
}

@article{press_harvey_2020,
  title        = {Harvey Weinstein trial hears from expert on unreliable memories},
  journal      = {The Guardian},
  year         = {2020},
  month        = {Feb},
  day          = {7},
  url          = {https://www.theguardian.com/film/2020/feb/07/harvey-weinstein-trial-unreliable-memories-elizabeth-loftus},
}

@article{fahimipour_wild_2023,
	title = {Wild animals suppress the spread of socially transmitted misinformation},
	volume = {120},
	url = {https://www.pnas.org/doi/10.1073/pnas.2215428120},
	doi = {10.1073/pnas.2215428120},
	number = {14},
	urldate = {2025-12-11},
	journal = {Proceedings of the National Academy of Sciences},
	author = {Fahimipour, Ashkaan K. and Gil, Michael A. and Celis, Maria Rosa and Hein, Gabriel F. and Martin, Benjamin T. and Hein, Andrew M.},
	month = apr,
	year = {2023},
	note = {Publisher: Proceedings of the National Academy of Sciences},
	pages = {e2215428120},
}

@article{stetka_have_2025,
	title = {Have people ‘had enough of experts’? {The} impact of populism and pandemic misinformation on institutional trust in comparative perspective},
	volume = {28},
	issn = {1369-118X},
	shorttitle = {Have people ‘had enough of experts’?},
	url = {https://doi.org/10.1080/1369118X.2024.2413121},
	doi = {10.1080/1369118X.2024.2413121},
	number = {6},
	urldate = {2025-12-11},
	journal = {Information, Communication \& Society},
	author = {Štětka, Václav and Brandao, Francisco and Mihelj, Sabina and Tóth, Fanni and Hallin, Daniel and Rothberg, Danilo and Ferracioli, Paulo and Klimkiewicz, Beata},
	month = apr,
	year = {2025},
	note = {Publisher: Routledge
\_eprint: https://doi.org/10.1080/1369118X.2024.2413121},
	pages = {1039--1060},
}

@incollection{akyuz_impact_2021,
	title = {The {Impact} of {Misinformation} and {Preferences} of {News} {Sources} on {Institutional} {Trust} {Perception} in the {COVID}-19 {Process}},
	copyright = {Access limited to members},
	isbn = {978-1-7998-7164-4},
	url = {https://www.igi-global.com/chapter/the-impact-of-misinformation-and-preferences-of-news-sources-on-institutional-trust-perception-in-the-covid-19-process/www.igi-global.com/chapter/the-impact-of-misinformation-and-preferences-of-news-sources-on-institutional-trust-perception-in-the-covid-19-process/278938},
	
	urldate = {2025-12-11},
	booktitle = {Impact of {Infodemic} on {Organizational} {Performance}},
	publisher = {IGI Global Scientific Publishing},
	author = {Akyüz, Selman Selim},
	year = {2021},
	doi = {10.4018/978-1-7998-7164-4.ch017},
	pages = {291--310},
}

@misc{hearst_qanon_2022,
	title = {{QAnon} and the {Rebirth} of the {Satanic} {Panic} in the {Digital} {Age}},
	author = {Hearst, Megan},
	year = {2022},
	note = {Publisher: Communication, Culture \& Technology},
}

@inproceedings{nguyen_sources_2012,
	title = {Sources of misinformation in {Online} {Social} {Networks}: {Who} to suspect?},
	doi = {10.1109/MILCOM.2012.6415780},
	booktitle = {{MILCOM} 2012 - 2012 {IEEE} {Military} {Communications} {Conference}},
	author = {Nguyen, Dung T. and Nguyen, Nam P. and Thai, My T.},
	year = {2012},
	pages = {1--6},
}

@book{kuhn_structure_1962,
	address = {Chicago, IL},
	edition = {1st},
	title = {The {Structure} of {Scientific} {Revolutions}},
	publisher = {University of Chicago Press},
	author = {Kuhn, Thomas S.},
	year = {1962},
}

@misc{posetti_short_2018, 
	title = {A {Short} {Guide} to the {History} of ‘{Fake} {News}’ and {Disinformation}},
	author = {Posetti, Julie and Matthews, Alice},
	month = jul,
	year = {2018},
	note = {Published: PDF file, International Center for Journalists (ICFJ)},
}

@article{donovan_express_2025,
	title = {{EXPRESS}: {A} {Short} {History} of {Misinformation}-at-scale and {Efforts} to {Mitigate} {It}},
	volume = {0},
	url = {https://doi.org/10.1177/07439156251384249},
	doi = {10.1177/07439156251384249},
	number = {ja},
	journal = {Journal of Public Policy and Marketing},
	author = {Donovan, Joan},
	year = {2025},
}

@article{loftus_reconstruction_1974,
	title = {Reconstruction of automobile destruction: {An} example of the interaction between language and memory},
	volume = {13},
	issn = {0022-5371},
	shorttitle = {Reconstruction of automobile destruction},
	url = {https://www.sciencedirect.com/science/article/pii/S0022537174800113},
	doi = {10.1016/S0022-5371(74)80011-3},
	number = {5},
	urldate = {2025-12-04},
	journal = {Journal of Verbal Learning and Verbal Behavior},
	author = {Loftus, Elizabeth F. and Palmer, John C.},
	month = oct,
	year = {1974},
	pages = {585--589},
}

@article{loftus_misinformation_2024,
	title = {Misinformation – past, present, and future},
	volume = {30},
	issn = {1068-316X},
	url = {https://doi.org/10.1080/1068316X.2023.2219813},
	doi = {10.1080/1068316X.2023.2219813},
	number = {4},
	urldate = {2025-12-04},
	journal = {Psychology, Crime \& Law},
	author = {Loftus, Elizabeth F. and Klemfuss, J. Zoe},
	month = apr,
	year = {2024},
	pages = {312--318},
}

@article{otgaar_protecting_2020,
	title = {Protecting {Against} {Misinformation}: {Examining} the {Effect} of {Empirically} {Based} {Investigative} {Interviewing} on {Misinformation} {Reporting}},
	volume = {36},
	url = {https://link.springer.com/article/10.1007/s11896-020-09401-2},
	doi = {10.1007/s11896-020-09401-2},
	number = {4},
	journal = {Journal of Police and Criminal Psychology},
	author = {Otgaar, Henry and de Ruiter, Corine and Sumampouw, Nathanael and Erens, Brenda and Muris, Peter},
	year = {2020},
	pages = {758--768},
}

@article{news_ghislaine_2021,
	title = {‘{Ghislaine} {Maxwell} trial: “{False} memory” expert testifies for defence’},
	url = {https://www.bbc.com/news/world-us-canada-59688787},
	journal = {BBC News},
	author = {News, B. B. C.},
	month = dec,
	year = {2021},
}

@article{faedda_fake_2024,
	title = {Fake memories: {A} meta-analysis on the effect of fake news on the creation of false memories and false beliefs},
	volume = {3},
	url = {https://doi.org/10.1017/mem.2024.14},
	doi = {10.1017/mem.2024.14},
	journal = {Memory, Mind \& Media},
	author = {Faedda, Sara and Marengo, Davide and Coscia, Matteo},
	year = {2024},
	pages = {e17},
}

@article{grady_primary_2023,
	title = {From {Primary} to {Presidency}: {Fake} {News}, {False} {Memory}, and {Changing} {Attitudes} in the 2016 {Election}},
	volume = {11},
	url = {https://doi.org/10.5964/jspp.10203},
	doi = {10.5964/jspp.10203},
	number = {1},
	journal = {Journal of Social and Political Psychology},
	author = {Grady, Rebecca Hofstein and Ditto, Peter H. and Loftus, Elizabeth F. and Levine, Linda J. and Greenspan, Rachel L. and Relihan, Daniel P.},
	year = {2023},
	pages = {6--24},
}

@article{marsh_how_2012,
	title = {How the {Truth} {Gets} {Twisted}},
	url = {https://stanfordmag.org/contents/how-the-truth-gets-twisted},
	journal = {Stanford Magazine},
	author = {Marsh, Ann and Lorge, Greta},
	month = dec,
	year = {2012},
}

@article{feeld_qanon_2025,
	title = {{QAnon} {Hasn}’t {Disappeared}. {It}’s in {America}’s {Bloodstream}.},
	url = {https://jacobin.com/2025/02/qanon-legacy-conspiracy-trump-patel},
	journal = {Jacobin},
	author = {Feeld, Julian},
	month = feb,
	year = {2025},
}

@article{alshaabi2021storywrangler,
  title        = {Storywrangler: A massive exploratorium for sociolinguistic, cultural, socioeconomic, and political timelines using Twitter},
  author       = {Thayer Alshaabi and Jane L. Adams and Michael V. Arnold and Joshua R. Minot and David R. Dewhurst and Andrew J. Reagan and Christopher M. Danforth and Peter Sheridan Dodds},
  journal      = {Science Advances},
  volume       = {7},
  number       = {29},
  pages        = {eabe6534},
  year         = {2021},
  doi          = {10.1126/sciadv.abe6534},
  url          = {https://advances.sciencemag.org/content/7/29/eabe6534}
}

@article{brashierAgingEraFake2020,
  title = {Aging in an {{Era}} of {{Fake News}}},
  author = {Brashier, Nadia M. and Schacter, Daniel L.},
  year = 2020,
  month = jun,
  journal = {Current directions in psychological science},
  volume = {29},
  number = {3},
  pages = {316--323},
  issn = {0963-7214},
  doi = {10.1177/0963721420915872},
  urldate = {2025-12-15},
  pmcid = {PMC7505057},
  pmid = {32968336}
}

@article{gradyNeverthelessPartisanshipPersisted2021,
  title = {Nevertheless, Partisanship Persisted: Fake News Warnings Help Briefly, but Bias Returns with Time},
  shorttitle = {Nevertheless, Partisanship Persisted},
  author = {Grady, Rebecca Hofstein and Ditto, Peter H. and Loftus, Elizabeth F.},
  year = 2021,
  month = jul,
  journal = {Cognitive Research: Principles and Implications},
  volume = {6},
  number = {1},
  pages = {52},
  issn = {2365-7464},
  doi = {10.1186/s41235-021-00315-z},
  urldate = {2025-12-15},
  langid = {english}
}

@article{murphyFalseMemoriesFake2019,
  title = {False {{Memories}} for {{Fake News During Ireland}}'s {{Abortion Referendum}}},
  author = {Murphy, Gillian and Loftus, Elizabeth F. and Grady, Rebecca Hofstein and Levine, Linda J. and Greene, Ciara M.},
  year = 2019,
  month = oct,
  journal = {Psychological Science},
  volume = {30},
  number = {10},
  pages = {1449--1459},
  issn = {1467-9280},
  doi = {10.1177/0956797619864887},
  langid = {english},
  pmid = {31432746}
}

@article{HUENEMANN1956623,
  title = {Combating Food Misinformation and Quackery},
  author = {Huenemann, Ruth L.},
  date = {1956},
  journaltitle = {Journal of the American Dietetic Association},
  volume = {32},
  number = {7},
  year = 1956,
  journal = {Journal of the American Dietetic Association},
  pages = {623--626},
  issn = {0002-8223},
  doi = {10.1016/S0002-8223(21)16192-9},
  url = {https://www.sciencedirect.com/science/article/pii/S0002822321161929}
}

@article{johnInformationMisinformationGained1957,
  title = {Information and Misinformation Gained from Fasting Blood Sugar Alone in Diabetes Therapy.},
  author = {John, H.J.},
  year = 1957,
  journal = {Ohio medicine: Journal of the Ohio State Medical Association},
  volume = {53},
  number = {11},
  pages = {1284--1287},
  issn = {0892-2454},
  langid = {english}
}

@article{gindinModelSovietFirm1970,
  title = {A Model of the {{Soviet}} Firm},
  author = {Gindin, Sam},
  date = {1970},
  year = 1970,  
  journal = {Economics of Planning},
  shortjournal = {Econ Plann},
  volume = {10},
  number = {3},
  pages = {145--157},
  publisher = {Kluwer Academic Publishers},
  issn = {1574-0277},
  doi = {10.1007/BF02342947},
  langid = {english}
}

@inproceedings{michelCategoricalApproachDistributed1989,
  title = {Categorical Approach to Distributed Systems, Expressibility and Knowledge},
  booktitle = {Proc {{Annu ACM Symp Princ Distrib Comput}}},
  author = {Michel, Ruben},
  year = 1989,
  pages = {129--143},
  publisher = {Publ by ACM},
  isbn = {978-0-89791-326-3},
  langid = {english}
}

@article{loftusPlantingMisinformationHuman2005b,
  title = {Planting Misinformation in the Human Mind: A 30-Year Investigation of the Malleability of Memory},
  shorttitle = {Planting Misinformation in the Human Mind},
  author = {Loftus, Elizabeth F.},
  year = 2005,
  journal = {Learning \& Memory (Cold Spring Harbor, N.Y.)},
  volume = {12},
  number = {4},
  pages = {361--366},
  issn = {1072-0502},
  doi = {10.1101/lm.94705},
  langid = {english},
  pmid = {16027179}
}

@article{ohEXPLORATIONSOCIALMEDIA2010,
  title = {An exploration of social media in extreme events: Rumor theory and Twitter during the Haiti earthquake},
  shorttitle = {{{AN EXPLORATION OF SOCIAL MEDIA IN EXTREME EVENTS}}},
  author = {Oh, Onook and Kwon, Kyounghee and Rao, H.},
  year = 2010,
  month = jan,
  journal = {ICIS 2010 Proceedings}
}

@article{eysenbachInfodemiologyEpidemiologyMisinformation2002,
  title = {Infodemiology: The Epidemiology of (Mis)Information},
  shorttitle = {Infodemiology},
  author = {Eysenbach, Gunther},
  year = 2002,
  month = dec,
  journal = {The American Journal of Medicine},
  volume = {113},
  number = {9},
  pages = {763--765},
  publisher = {Elsevier},
  issn = {0002-9343, 1555-7162},
  doi = {10.1016/S0002-9343(02)01473-0},
  urldate = {2025-12-15},
  langid = {english}
}

@article{suttonBackchannelsFrontLines2008,
  title = {Backchannels on the {{Front Lines}}: {{Emergent Uses}} of {{Social Media}} in the 2007 {{Southern California Wildfires}}},
  author = {Sutton, Jeannette and Palen, Leysia and Shklovski, Irina},
  year = 2008,
  langid = {english},
  journal = {Proceedings of ISCRAM 2008 - 5th International Conference on Information Systems for Crisis Response and Management}
}

@article{chewPandemicsAgeTwitter2010,
  title = {Pandemics in the Age of {{Twitter}}: {{Content}} Analysis of Tweets during the 2009 {{H1N1}} Outbreak},
  shorttitle = {Pandemics in the Age of {{Twitter}}},
  author = {Chew, Cynthia and Eysenbach, Gunther},
  year = 2010,
  journal = {PLoS ONE},
  volume = {5},
  number = {11},
  issn = {1932-6203},
  doi = {10.1371/journal.pone.0014118},
  langid = {english}
}

@article{lewandowskyMisinformationUnderstandingCoping2017,
  title = {Beyond Misinformation: {{Understanding}} and Coping with the “Post-Truth” Era.},
  shorttitle = {Beyond Misinformation},
  author = {Lewandowsky, Stephan and Ecker, Ullrich K. H. and Cook, John},
  date = {2017-12},
  journaltitle = {Journal of Applied Research in Memory and Cognition},
  journal = {Journal of Applied Research in Memory and Cognition},
  volume = {6},
  number = {4},
  pages = {353--369},
  issn = {2211-369X, 2211-3681},
  doi = {10.1016/j.jarmac.2017.07.008},
  url = {https://doi.apa.org/doi/10.1016/j.jarmac.2017.07.008},
  urldate = {2025-12-16},
  langid = {english}
}

@article{hymanMisinformationWorldviewsPosttruth2017,
  title = {Misinformation and Worldviews in the Post-Truth Information Age: {{Commentary}} on {{Lewandowsky}}, {{Ecker}}, and {{Cook}}.},
  shorttitle = {Misinformation and Worldviews in the Post-Truth Information Age},
  author = {Hyman, Ira E. and Jalbert, Madeline C.},
  date = {2017-12},
  journal = {Journal of Applied Research in Memory and Cognition},
  shortjournal = {Journal of Applied Research in Memory and Cognition},
  volume = {6},
  number = {4},
  pages = {377--381},
  issn = {2211-369X, 2211-3681},
  doi = {10.1016/j.jarmac.2017.09.009},
  url = {https://doi.apa.org/doi/10.1016/j.jarmac.2017.09.009},
  urldate = {2025-12-15},
  langid = {english}
}

@article{kongBriefNaturalHistory2025,
  title = {A Brief Natural History of Misinformation},
  author = {Kong, Ling-Wei and Gallart, Lucas and Grassick, Abigail G. and Love, Jay W. and Nayak, Amlan and Hein, Andrew M.},
  date = {2025-12-10},
  journal = {Journal of The Royal Society Interface},
  shortjournal = {J R Soc Interface},
  year = 2025,
  volume = {22},
  number = {233},
  pages = {20250161},
  issn = {1742-5689},
  doi = {10.1098/rsif.2025.0161},
  url = {https://doi.org/10.1098/rsif.2025.0161},
  urldate = {2025-12-17}
}

@online{arnoldMisinformationInevitableBiological,
  title = {Misinformation Is an Inevitable Biological Reality across Nature, Researchers Argue},
  author = {Arnold, Paul},
  url = {https://phys.org/news/2025-12-misinformation-inevitable-biological-reality-nature.html},
  urldate = {2025-12-17},
  year = 2025,
  langid = {english}
}

@thesis{poudelCuratedRealitiesIntersection2025,
  type = {thesis},
  title = {Curated {{Realities}}: {{The Intersection}} of {{Algorithmic Curation}}, {{Bias}}, and {{Framing}} in {{Digital Media}}},
  shorttitle = {Curated {{Realities}}},
  author = {Poudel, Amrit},
  date = {2025-12-09},
  year = 2025,
  institution = {University of Notre Dame},
  doi = {10.7274/30729425.v1},
  url = {https://curate.nd.edu/articles/thesis/Curated_Realities_The_Intersection_of_Algorithmic_Curation_Bias_and_Framing_in_Digital_Media/30729425/1},
  urldate = {2025-12-18},
  langid = {english}
}

@book{chomsky1988manufacturing,
  title     = {Manufacturing Consent: The Political Economy of the Mass Media},
  author    = {Chomsky, Noam and Herman, Edward S.},
  year      = {1988},
  publisher = {Pantheon Books},
  address   = {New York}
}

@book{williamsTelevisionTechnologyCultural2004,
  title = {Television: {{Technology}} and {{Cultural Form}}},
  shorttitle = {Television},
  author = {Williams, Raymond},
  date = {2004-05-31},
  edition = {3},
  publisher = {Routledge},
  location = {London},
  doi = {10.4324/9780203426647},
  isbn = {978-0-203-42664-7},
  pagetotal = {192}
}

@book{mcluhan_understanding_media_1964,
  author    = {McLuhan, Marshall},
  title     = {Understanding Media: The Extensions of Man},
  publisher = {McGraw-Hill / New American Library},
  address   = {New York},
  year      = {1964},
  isbn      = {0070454361},
}

@article{cinelliCOVID19SocialMedia2020,
  title = {The {{COVID-19}} Social Media Infodemic},
  author = {Cinelli, Matteo and Quattrociocchi, Walter and Galeazzi, Alessandro and Valensise, Carlo Michele and Brugnoli, Emanuele and Schmidt, Ana Lucia and Zola, Paola and Zollo, Fabiana and Scala, Antonio},
  date = {2020-10-06},
  year = 2020,
  journal = {Scientific Reports},
  shortjournal = {Sci Rep},
  volume = {10},
  eprint = {33024152},
  eprinttype = {pubmed},
  pages = {16598},
  issn = {2045-2322},
  doi = {10.1038/s41598-020-73510-5},
  url = {https://pmc.ncbi.nlm.nih.gov/articles/PMC7538912/},
  urldate = {2025-12-20},
  pmcid = {PMC7538912}
}

@book{eichhornContent2022,
  title = {Content},
  author = {Eichhorn, Kate},
  date = {2022-05-10},
  year = 2022,
  series = {The {{MIT Press Essential Knowledge}} Series},
  publisher = {MIT Press},
  location = {Cambridge, MA, USA},
  isbn = {978-0-262-54328-6},
  langid = {english},
  pagetotal = {192}
}

@article{bovetInfluenceFakeNews2019,
  title = {Influence of Fake News in {{Twitter}} during the 2016 {{US}} Presidential Election},
  author = {Bovet, Alexandre and Makse, Hern{\'a}n A.},
  year = 2019,
  month = jan,
  journal = {Nature Communications},
  volume = {10},
  number = {1},
  pages = {7},
  publisher = {Nature Publishing Group},
  issn = {2041-1723},
  doi = {10.1038/s41467-018-07761-2},
  urldate = {2025-12-29},
  copyright = {2019 The Author(s)},
  langid = {english}
}

@article{flaminoPoliticalPolarizationNews2023,
  title = {Political Polarization of News Media and Influencers on {{Twitter}} in the 2016 and 2020 {{US}} Presidential Elections},
  author = {Flamino, James and Galeazzi, Alessandro and Feldman, Stuart and Macy, Michael W. and Cross, Brendan and Zhou, Zhenkun and Serafino, Matteo and Bovet, Alexandre and Makse, Hernán A. and Szymanski, Boleslaw K.},
  date = {2023-06},
  year=2023,  
  journal = {Nature Human Behaviour},
  shortjournal = {Nat Hum Behav},
  volume = {7},
  number = {6},
  eprint = {36914806},
  eprinttype = {pubmed},
  pages = {904--916},
  issn = {2397-3374},
  doi = {10.1038/s41562-023-01550-8},
  langid = {english},
  pmcid = {PMC10289895}
}

@online{hanMeasuringCoevolutionOnline2025,
  title = {Measuring the Co-Evolution of Online Engagement with (Mis)Information and Its Visibility at Scale},
  author = {Han, Yueting and Turrini, Paolo and Bazzi, Marya and Andrighetto, Giulia and Polizzi, Eugenia and Domenico, Manlio De},
  date = {2025-06-06},
  eprint = {2506.06106},
  eprinttype = {arXiv},
  eprintclass = {cs},
  doi = {10.48550/arXiv.2506.06106},
  url = {http://arxiv.org/abs/2506.06106},
  urldate = {2025-12-29},
  year=2025,
  langid = {english},
  pubstate = {prepublished}
}

@book{tremblay_pepin_intrigues_2025,
  author    = {Simon Tremblay-Pépin},
  title     = {Intrigues: Petit manuel pour une critique des relations publiques},
  editor    = {Lux Éditeur},
  year      = {2025},
  address   = {Montréal, Québec, Canada},
  isbn      = {978-2-89833-208-1},
  note      = {French},
}

@article{Allen2020FakeNews,
  title        = {Evaluating the fake news problem at the scale of the information ecosystem},
  author       = {Allen, Jennifer and Howland, Baird and Mobius, Markus and Rothschild, David and Watts, Duncan J.},
  journal      = {Science Advances},
  year         = {2020},
  volume       = {6},
  number       = {14},
  pages        = {eaay3539},
  doi          = {10.1126/sciadv.aay3539},
  url          = {https://www.science.org/doi/10.1126/sciadv.aay3539}
}

@article{konesPreventionFantasyFuture2011,
  title = {Is Prevention a Fantasy, or the Future of Medicine? {{A}} Panoramic View of Recent Data, Status, and Direction in Cardiovascular Prevention},
  shorttitle = {Is Prevention a Fantasy, or the Future of Medicine?},
  author = {Kones, Richard},
  date = {2011-02},
  year = 2011,
  journaltitle = {Therapeutic Advances in Cardiovascular Disease},
  journal = {Therapeutic Advances in Cardiovascular Disease},  
  shortjournal = {Ther Adv Cardiovasc Dis},
  volume = {5},
  number = {1},
  eprint = {21183531},
  eprinttype = {pubmed},
  pages = {61--81},
  issn = {1753-9455},
  doi = {10.1177/1753944710391350},
  langid = {english}
}

@incollection{Zagzebski2012,
  author       = {Linda Trinkaus Zagzebski},
  title        = {Introduction},
  booktitle    = {Epistemic Authority: A Theory of Trust, Authority, and Autonomy in Belief},
  year         = {2012},
  publisher    = {Oxford University Press},
  address      = {Oxford},
  pages        = {1--3},
  doi          = {10.1093/acprof:oso/9780199936472.003.0001},
  url          = {https://doi.org/10.1093/acprof:oso/9780199936472.003.0001}
}

@book{gierynCulturalBoundariesScience1999,
  title = {Cultural {{Boundaries}} of {{Science}}: {{Credibility}} on the {{Line}}},
  shorttitle = {Cultural {{Boundaries}} of {{Science}}},
  author = {Gieryn, Thomas F.},
  year = 1999,
  month = jan,
  publisher = {University of Chicago Press},
  address = {Chicago, IL},
  urldate = {2026-01-03},
  isbn = {978-0-226-29262-5},
  langid = {english}
}

@article{priceDoomscrollingCOVID19Negative2022,
  title = {Doomscrolling during {{COVID-19}}: {{The}} Negative Association between Daily Social and Traditional Media Consumption and Mental Health Symptoms during the {{COVID-19}} Pandemic},
  shorttitle = {Doomscrolling during {{COVID-19}}},
  author = {Price, Matthew and Legrand, Alison C. and Brier, Zoe M. F. and {van Stolk-Cooke}, Katherine and Peck, Kelly and Dodds, Peter Sheridan and Danforth, Christopher M. and Adams, Zachary W.},
  year = 2022,
  journal = {Psychological Trauma: Theory, Research, Practice, and Policy},
  volume = {14},
  number = {8},
  pages = {1338--1346},
  publisher = {Educational Publishing Foundation},
  address = {US},
  issn = {1942-969X},
  doi = {10.1037/tra0001202}
}

@article{blondel2008fast,
  title     = {Fast unfolding of communities in large networks},
  author    = {Blondel, Vincent D. and Guillaume, Jean-Loup and Lambiotte, Renaud and Lefebvre, Etienne},
  journal   = {Journal of Statistical Mechanics: Theory and Experiment},
  volume    = {2008},
  number    = {10},
  pages     = {P10008},
  year      = {2008},
  publisher = {IOP Publishing},
  doi       = {10.1088/1742-5468/2008/10/P10008}
}

@misc{scopus,
  title        = {Scopus},
  author       = {{Elsevier B.V.}},
  howpublished = {\url{https://www.scopus.com}},
  year         = {2025},
  note         = {Bibliographic data retrieved January 8, 2026}
}

@book{ellul1973propaganda,
  title = {Propaganda: The Formation of Men's Attitudes},
  author = {Ellul, Jacques},
  translator = {Kellen, Konrad},
  year = {1973},
  publisher = {Vintage Books},
  address = {New York},
  isbn = {9780394718743}
}

@misc{linderMcMartinPreschoolAbuse2007,
  type = {{{SSRN Scholarly Paper}}},
  title = {The {{McMartin Preschool Abuse Trial}}},
  author = {Linder, Douglas},
  year = 2007,
  number = {1030559},
  eprint = {1030559},
  publisher = {Social Science Research Network},
  address = {Rochester, NY},
  doi = {10.2139/ssrn.1030559},
  urldate = {2026-01-19},
  archiveprefix = {Social Science Research Network},
  langid = {english}
}

@article{Reinhold1990LongestTrial,
  author       = {Robert Reinhold},
  title        = {The Longest Trial\,---\,A Post-Mortem; Collapse of Child-Abuse Case: So Much Agony for So Little},
  journal      = {The New York Times},
  date         = {1990-01-24},
  url          = {https://www.nytimes.com/1990/01/24/us/longest-trial-post-mortem-collapse-child-abuse-case-so-much-agony-for-so-little.html},
  note         = {Accessed via New York Times archive},
}

@article{Schreiber2006Suggestive,
  author       = {Nadja Schreiber and Lisa D. Bellah and Yolanda Martinez and Kristin A. McLaurin and Renata Strok and Sena Garven and James M. Wood},
  title        = {Suggestive interviewing in the McMartin Preschool and Kelly Michaels daycare abuse cases: A case study},
  journal      = {Social Influence},
  year         = {2006},
  volume       = {1},
  number       = {1},
  pages        = {16--47},
  doi          = {10.1080/15534510500361739},
  url          = {https://scholarworks.utep.edu/cgi/viewcontent.cgi?article=1014&context=james_wood},
}

@article{SuburbanUncanny2019,
  title = {The {{Suburban Uncanny}}},
  author = {Colin Dickey},
  year = 2019,
  month = jun,
  journal = {Los Angeles Review of Books},
  urldate = {2026-01-19},
  howpublished = {https://lareviewofbooks.org/article/the-suburban-uncanny}
}

@article{shaw2026,
  title = {Where Was Skepticism in {{Media}}? : {{Pack}} Journalism and Hysteria Marked Early Coverage of the {{McMartin}} Case. {{Few}} Journalists Stopped to Question the Believability of the Prosecution's Charges. - {{Los Angeles Times}}},
  author={David Shaw},
  journal = {The Los Angeles Times},
  urldate = {2026-01-19},
  howpublished = {https://www.latimes.com/archives/la-xpm-1990-01-19-mn-226-story.html}
}

@misc{WildClaimsMass2024,
  title = {Wild Claims of Mass Child Molestation Rocked an {{L}}.{{A}}. Beach Town. {{Truth}} Was the First Casualty},
  year = 2024,
  author = {Christopher Goffard},
  month = jul,
  journal = {Los Angeles Times},
  urldate = {2026-01-19},
  chapter = {California},
  howpublished = {https://www.latimes.com/california/story/2024-07-17/crimes-of-the-times-mcmartin-preschool},
  langid = {american}
}

@article{bentonEverythingOldAgain2025,
  title = {Everything {{Old}} Is {{Q Again}}: {{An Analysis}} of the {{Resurgent}} \#{{PIZZAGATE Myth}} in {{Social Media}}},
  shorttitle = {Everything {{Old}} Is {{Q Again}}},
  author = {Benton, Bond and {Peterka-Benton}, Daniela},
  year = 2025,
  month = jan,
  journal = {Popular Culture Studies Journal}
}

@book{emerson2024unmask,
  title     = {Unmask Alice: LSD, Satanic Panic, and the Imposter Behind the World's Most Notorious Diaries},
  author    = {Emerson, Rick},
  publisher = {BenBella Books},
  year      = {2024},
  isbn      = {9781637745182},
  url       = {https://www.usu.edu/honors/book-labs/2025/summer/unmask-alice},
  note      = {Discussed in a 2025 Utah State University Honors Book Lab (Summer 2025) by Katie Luder},
}

@article{otgaarReturnRepressedPersistent2019,
  title = {The {{Return}} of the {{Repressed}}: {{The Persistent}} and {{Problematic Claims}} of {{Long-Forgotten Trauma}}},
  shorttitle = {The {{Return}} of the {{Repressed}}},
  author = {Otgaar, Henry and Howe, Mark L. and Patihis, Lawrence and Merckelbach, Harald and Lynn, Steven Jay and Lilienfeld, Scott O. and Loftus, Elizabeth F.},
  year = 2019,
  month = nov,
  journal = {Perspectives on Psychological Science},
  volume = {14},
  number = {6},
  pages = {1072--1095},
  publisher = {SAGE Publications Inc},
  issn = {1745-6916},
  doi = {10.1177/1745691619862306},
  urldate = {2026-01-19},
  langid = {english}
}

@article{guessExposureUntrustworthyWebsites2020,
  title = {Exposure to Untrustworthy Websites in the 2016 {{US}} Election},
  author = {Guess, Andrew M. and Nyhan, Brendan and Reifler, Jason},
  year = 2020,
  month = may,
  journal = {Nature Human Behaviour},
  volume = {4},
  number = {5},
  pages = {472--480},
  publisher = {Nature Publishing Group},
  issn = {2397-3374},
  doi = {10.1038/s41562-020-0833-x},
  urldate = {2026-01-26},
  copyright = {2020 The Author(s), under exclusive licence to Springer Nature Limited},
  langid = {english}
}

@article{scharlachVALUEAFFORDANCESSOCIAL2023,
  title = {The value affordances of social media engagement features},
  author = {Scharlach, Rebecca and Hallinan, Blake},
  year = 2023,
  month = dec,
  journal = {AoIR Selected Papers of Internet Research},
  issn = {2162-3317},
  doi = {10.5210/spir.v2023i0.13491},
  urldate = {2026-02-06},
  copyright = {Copyright (c) 2023 AoIR Selected Papers of Internet Research},
  langid = {english}
}

@article{norburyMisinformationTacticsProtect2021,
  title = {Misinformation Tactics Protect Rare Birds from Problem Predators},
  author = {Norbury, Grant L. and Price, Catherine J. and Latham, M. Cecilia and Brown, Samantha J. and Latham, A. David M. and Brownstein, Gretchen E. and Ricardo, Hayley C. and McArthur, Nikki J. and Banks, Peter B.},
  year = 2021,
  month = mar,
  journal = {Science Advances},
  volume = {7},
  number = {11},
  pages = {eabe4164},
  publisher = {American Association for the Advancement of Science},
  doi = {10.1126/sciadv.abe4164},
  urldate = {2026-02-25}
}
\end{document}